\documentclass[12pt, a4paper]{article}

\usepackage{color,amsfonts,amsmath,nicefrac,graphicx,booktabs,comment,natbib}
\usepackage[normalem]{ulem}

\def\bSig\mathbf{\Sigma}

\newcommand{\bs}[1]{\boldsymbol{#1}}
\newcommand{\asym}{\overset{a}{\sim}}
\newcommand{\tendist}{\overset{d}{\longrightarrow}}

\begin{document}

\title{Conflict diagnostics for evidence synthesis in a multiple testing framework}

\author{Anne M. Presanis, David Ohlssen, Kai Cui, \\ Magdalena Rosinska,
  Daniela De Angelis}
\date{\today}

 \maketitle
 \begin{center}
   \emph{Medical Research Council Biostatistics Unit, University of
     Cambridge, U.K. \\ Novartis Pharmaceuticals Corporation, East
     Hanover, NJ, U.S.A. \\ Department of Epidemiology, National
     Institute of Public Health, \\
National Institute of Hygiene, Warsaw, Poland} \\ \bigskip
   e-mail: anne.presanis@mrc-bsu.cam.ac.uk
 \end{center}

\begin{abstract}\noindent
Evidence synthesis models that combine multiple datasets of varying design, to estimate quantities that cannot be directly observed, require the formulation of complex probabilistic models that can be expressed as graphical models. An assessment of whether the different datasets synthesised contribute information that is consistent with each other, and in a Bayesian context, with the prior distribution, is a crucial component of the model criticism process. However, a systematic assessment of conflict suffers from the multiple testing problem, through testing for conflict at multiple locations in a model. We demonstrate the systematic use of conflict diagnostics, while accounting for the multiple hypothesis tests of no conflict at each location in the graphical model. The method is illustrated by a network meta-analysis to estimate treatment effects in smoking cessation programs and an evidence synthesis to estimate HIV prevalence in Poland.
\bigskip

\end{abstract}

\section{Introduction \label{sec_intro}}
Evidence synthesis refers to the use of complex statistical models
that combine multiple, disparate and imperfect sources of evidence to
estimate quantities on which direct information is unavailable or
inadequate
\citep[e.g.][]{AdesSutton2006,WeltonEtAl2012,DeAngelisEtAl2014}. Such evidence synthesis models are typically graphical models represented by a directed acyclic graph (DAG) $\mathcal{G}(\bs{V}, \bs{E})$, where $\bs{V}$ and $\bs{E}$ are sets of nodes and edges respectively, encoding conditional independence assumptions \citep{Lauritzen1996}. With increased computational power, models of the form of
$\mathcal{G}(\bs{V}, \bs{E})$ have proliferated, requiring also the
development of model criticism tools adapted to the challenges of
evidence synthesis. In a Bayesian framework, any of the prior
distribution, the assumed form of the likelihood and structural and
functional assumptions may conflict with the observed data or with
each other. To assess the consistency of each of these components,
various mixed- or posterior-predictive checks have been proposed. In
particular, the ``conflict p-value''
\citep{MarshallSpiegelhalter2007,GasemyrNatvig2009,PresanisEtAl2013,Gasemyr2015}
is a diagnostic calculated by splitting $\mathcal{G}(\bs{V}, \bs{E})$ into two independent sub-graphs (``partitions'') at a particular ``separator''
node $\phi$, to measure the consistency of the information provided by
each partition about the node (a ``node-split''). \citet{GasemyrNatvig2009} and \citet{PresanisEtAl2013} demonstrate how the conflict p-value may be evaluated in different contexts, including both one- and two-sided hypothesis tests, and \citet{Gasemyr2015} demonstrates the uniformity of the conflict p-value in a wide range of models.

The conflict p-value may be used in a targeted manner, searching for
conflict at particular nodes in a DAG. However, in complex evidence
syntheses, often the location of potential conflict may be unclear. A
systematic assessment of conflict throughout a DAG is then required to
locate problem areas \citep[e.g.][]{KrahnEtAl2013}. Such systematic
assessment, however, suffers from the multiple testing problem, either
through testing for conflict at each node in $\mathcal{G}(\bs{V}, \bs{E})$ or through the separation of $\mathcal{G}(\bs{V}, \bs{E})$
into more than two partitions to simultaneously test for conflict
between each pair-wise partition. Here we account for these multiple tests
by adopting the general hypothesis testing framework of
\citet{HothornEtAl2008,BretzEtAl2011}, allowing for simultaneous
multiple hypotheses in a parametric setting. They propose different
possible tests to account for multiplicity: we concentrate here on maximum-T type tests.

In section \ref{sec_egintro}, we define evidence synthesis before introducing the particular models that motivate our work on systematic conflict assessment: a network meta-analysis and a model for estimating HIV prevalence. Section \ref{sec_methods} describes the methods we use to test for conflict and account for the multiple tests we perform. We apply these methods to our examples in Section \ref{sec_applications} and end with a discussion in Section \ref{sec_discuss}.

\section{Motivating examples \label{sec_egintro}}
\begin{figure}
  \centering
  \includegraphics[width = \textwidth]{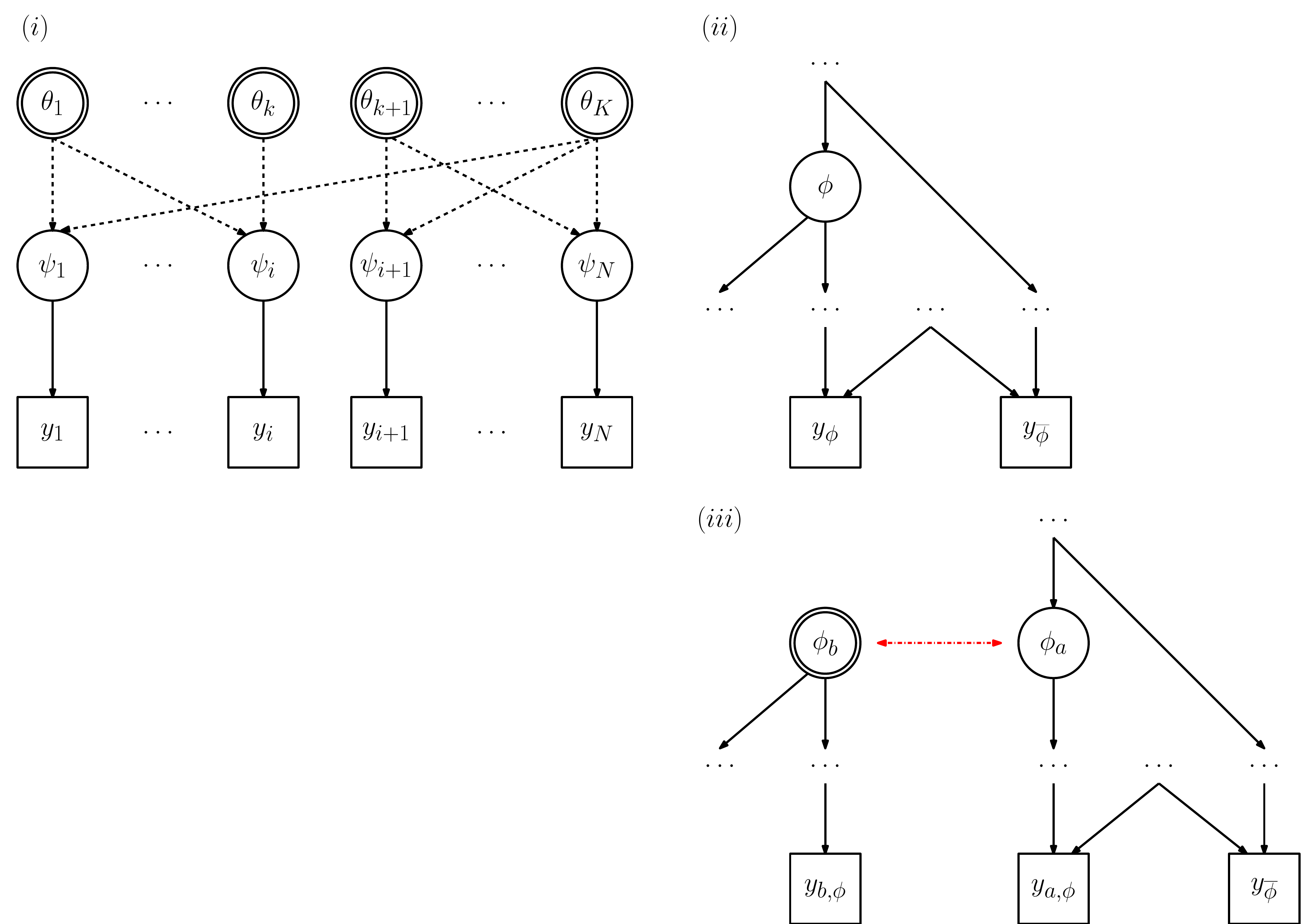}
  \caption{(i) Example DAG $\mathcal{G}(\bs{V}, \bs{E})$ showing a
    generic evidence synthesis. (ii) \& (iii) Example node-split at separator node $\phi$: (ii) original model $\mathcal{G}(\bs{\phi},\bs{y})$; (iii) node-split model. In (ii): the data $\bs{y} = \{\bs{y}_{\phi}, \bs{y}_{\overline{\phi}}\}$ comprise data $\bs{y}_{\phi}$ that are direct descendents of $\phi$; and the remaining data $\bs{y}_{\overline{\phi}}$. In (iii): when splitting $\mathcal{G}(\bs{\phi},\bs{y})$ into partitions $a$ and $b$, the data vector $\bs{y}_{\phi}$ is split into $\bs{y}_{a,\phi}$ and $\bs{y}_{b,\phi}$, whereas $\bs{y}_{\overline{\phi}}$ remains only in partition $a$. The partition $a$ data are therefore $\bs{y}_a = \{\bs{y}_{a,\phi}, \bs{y}_{\overline{\phi}}\}$ and the partition $b$ data are $\bs{y}_b = \bs{y}_{b,\phi}$. \label{fig_genSplit}}
\end{figure}

Formally, our goal is to estimate $K$ \emph{basic} parameters
$\bs{\theta} = (\theta_1, \ldots, \theta_K)$ given a collection of $N$
independent data sources $\bs{y} = (\bs{y}_1, \ldots, \bs{y}_N)$,
where each $\bs{y}_i, i \in 1, \ldots, N$ may be a vector or array of
data points. Each $\bs{y}_i$ provides information on a
\emph{functional} parameter $\psi_i$ (or potentially a vector of
functions $\bs{\psi}_i$). When $\psi_i = \theta_k$ is the identity
function, the data $\bs{y}_i$ are said to \emph{directly} inform
$\theta_k$. Otherwise, $\psi_i = \psi_i(\bs{\theta})$ is a function of
multiple parameters in $\bs{\theta}$: the $\bs{y}_i$ therefore provide \emph{indirect} information on these parameters. Given the
conditional independence of the datasets $\bs{y}_i$, the likelihood is
$L(\bs{\theta}; \bs{y}) = \prod_{i = 1}^NL_i(\psi_i(\bs{\theta});
\bs{y}_i)$, where $L_i(\psi_i(\bs{\theta}); \bs{y}_i)$ is the
likelihood contribution of $\bs{y}_i$ given the basic parameters
$\bs{\theta}$. In a Bayesian context, for a prior distribution
$p(\bs{\theta})$, the posterior distribution $p(\bs{\theta} \mid
\bs{y}) \propto p(\bs{\theta})L(\bs{\theta}; \bs{y})$ summarises all
information, direct and indirect, on $\bs{\theta}$. Let $\bs{\psi} =
(\psi_1, \ldots, \psi_N)$ be the set of functional parameters informed
by data and $\bs{\phi} = \{\bs{\theta}, \bs{\psi}\}$ be the set of all
unknown quantities, whether basic or functional. In this setup, the DAG $\mathcal{G}(\bs{V},\bs{E})$ representing the evidence synthesis model has a set of nodes $\bs{V} = \{\bs{\phi}, \bs{y}\}$ representing either known or unknown quantities; and the directed edges $\bs{E}$ represent dependencies between nodes. Each `child' node is independent of its `siblings'  conditional on their direct `parents'. The joint distribution of all nodes $\bs{V}$ is the product of the conditional distributions of each node given its direct parents. An example DAG of an evidence synthesis model is given in Figure \ref{fig_genSplit}(i). Circles denote unknown quantities: either basic parameters $\bs{\theta}$ that are `founder' nodes at the top of a DAG having a prior distribution (double circles); or functional parameters $\bs{\psi}$. Squares denote observed quantities, solid arrows represent stochastic distributional relationships, and dashed arrows represent deterministic functional relationships. This DAG could be extended to more complex hierarchical priors and models, where repetition over variables is represented by `plates', rounded rectangles around the repeated nodes, labelled by the range of repetition. In general, the set $\bs{V}$ may be larger than the set of basic and functional parameters, including also other intermediate nodes in the DAG, for example unit-level parameters in a hierarchical model. For brevity, from here on we will abbreviate any DAG to the notation $\mathcal{G}(\bs{\phi}, \bs{y})$.

\subsection{Network meta-analysis}

Network meta-analysis (NMA) is a specific type of evidence synthesis \citep{Salanti2012}, that generalises meta-analysis from the synthesis of studies measuring a treatment effect (e.g. of treatment B versus treatment A in a randomised clinical trial), to the synthesis of data on more than two treatment arms. The studies included in the NMA may not all measure the same treatment effects, but each study provides data on at least two of the treatments. For example, considering a set of treatments $\{A,B,C,D\}$, the network of trials may consist of studies of different ``designs'', i.e. with different subsets of the treatments included in each trial \citep{JacksonEtAl2014}, such as $\{ABC,ABD,BD,CD\}$. As with meta-analysis, NMA models can be implemented in either a two-stage or single-stage approach, as described more comprehensively elsewhere \citep{Salanti2012,JacksonEtAl2014}. Here we concentrate on a single-stage approach, where the original
data $Y_{di}^J$ for each treatment $J$ of study $i$ of design $d$ are available. A full likelihood model specifies 
$$
Y_{di}^{J} \sim f(p_{di}^{J} \mid w_{di}^{J})
$$ for some distribution $f(\cdot)$ and treatment outcome $p_{di}^{J}$ with associated information $w_{di}^{J}$. For example, if the data are numbers of events out of total numbers at risk of the event, then $w_{di}^{J}$ might be the denominator for treatment $J$. We might assume the data are realisations of a Binomial random variable, $Y_{di}^{J} \sim Bin(w_{di}^{J}, p_{di}^{J})$, 
where the proportion $p_{di}^{J}$ is a function of a study-specific baseline $\alpha_{di}$ representing a design/study-specific baseline treatment $B_d$ and a study-specific treatment contrast (log odds ratio) $\mu_{di}^{B_dJ}$, through a logistic model, $logit(p_{di}^J) = \alpha_{di} + \mu_{di}^{B_dJ}$.
The intercept is $\alpha_{di} = logit(p_{di}^{B_d})$. To complete the model specification requires parameterisation of the treatment effects $\mu_{di}^{AJ}$. A common effect model, for a network-wide reference treatment $A$, is given by
\begin{equation}
\mu_{di}^{AJ} = \eta^{AJ} \label{eqn_consFE}
\end{equation}
for each $J \neq A$, i.e. assumes that all studies of all designs measure the same treatment effects. The $\eta^{AJ}$ are basic parameters, of which there are the number of treatments in the network minus 1, representing the relative effectiveness of treatment $J$ compared to the network baseline treatment $A$. All other contrasts $\eta^{JK}, J,K \neq A$ are functional parameters, defined by assuming a set of \emph{consistency} equations $\eta^{JK} = \eta^{AK} - \eta^{AJ}$ for each $J,K \neq A$. These equations define a transitivity property of the treatment effects. The extension to a random-effects model, still under the consistency assumption, implies
\begin{equation}
\mu_{di}^{AJ} = \eta^{AJ} + \beta_{di}^{AJ} \label{eqn_consRE}
\end{equation}
where usually the random effects $\beta_{di}^{AJ}$, reflecting
between-study heterogeneity, are assumed normally distributed around
$0$, with a covariance structure defined as a square matrix
$\Sigma_{\beta}$ such that all entries on the leading diagonal are
$\sigma_{\beta}^2$ and all remaining entries are $\sigma_{\beta}^2/2$
\citep{Salanti2012,JacksonEtAl2015}. Figure \ref{fig_nmaDAGs} of the Supplementary Material shows the DAG structure of both the common and random effects models for a full likelihood setting where the outcome is binomial. The set of \emph{basic} parameters is denoted $\bs{\eta}_b = (\eta^{AJ})_{J \neq A}$ and the corresponding set of \emph{functional} parameters is denoted $\bs{\eta}_f = (\eta^{JK} = \eta^{AK} - \eta^{AJ})_{J,K \neq A}$. Note that the common-effect model is a special case of the random-effects model. In the Bayesian paradigm, we specify prior distributions for the basic parameters $\bs{\eta}_b$, the (nuisance) study-specific baselines $\alpha_{di}$, and in the case of the random treatment effects model, the common standard deviation parameter $\sigma_{\beta}$ in terms of which the variance-covariance matrix $\Sigma_{\beta}$ is defined. Note that any change in parameterisation of the model, for example changing treatment labels, will affect the joint prior distribution, making invariance challenging or even impossible in a Bayesian setting.

\paragraph{A smoking cessation example}
\citet{DiasEtAl2010}, amongst many others
\citep{LuAdes2006,HigginsEtAl2012,JacksonEtAl2015}, considered an NMA
of studies of smoking cessation. The network consists of 24 studies of
8 different designs, including 2 three-arm trials. Four smoking
cessation counselling programs are compared (Figure
\ref{fig_smkSpan}): A no intervention; B self-help; C individual counselling; D group counselling. The data (Supplementary Material Table \ref{tab_smkData}) are the number of individuals out of those
participating who have successfully ceased to smoke at 6-12 months
after enrollment. Here we fit the common- and random-effect models
under a consistency assumption and diffuse priors: Normal$(0,10^2)$ on the log-odds scale for
$\bs{\eta}_b$ and $\alpha_{di}$; and Uniform$(0,5)$ for
$\sigma_{\beta}$. We find (Supplementary Material Table \ref{tab_nmaConsistent}) that the deviance information criterion ($DIC$, \citet{SpiegelhalterEtAl2002}) prefers the random-effect model, suggesting it is necessary to explain the heterogeneity in the network. The estimates of the treatment effects from the random-effect model are
both somewhat different and more uncertain than those from the
common-effect model, agreeing with estimates found by others,
including \citet{DiasEtAl2010}. Moreover, the posterior expected
deviance for the random-effect model, $\mathbb{E}_{\theta \mid y}(D) =
54$, is slightly larger than the number of observations (50),
suggesting still some lack of fit to the data.

\paragraph{A single node-split model}
This residual lack of fit and the general potential in NMA for
variability between groups of direct and indirect information from
multiple studies that is excess to between-study heterogeneity
(``inconsistency'', \citet{LuAdes2006}) has motivated various
approaches to the detection and resolution of inconsistency
\citep{Lumley2002,LuAdes2006,DiasEtAl2010,HigginsEtAl2012,WhiteEtAl2012,JacksonEtAl2014}. \citet{DiasEtAl2010}
apply the idea of node-splitting, based on
\citet{MarshallSpiegelhalter2007}, to the NMA context, splitting a
single mean treatment effect $\eta^{JK}$ in the random effects
consistency model (\ref{eqn_consRE}). A DAG is partitioned into \emph{direct} evidence from studies directly comparing
$J$ and $K$ versus \emph{indirect} evidence from all remaining
studies. Specifically, for any study $i$ of design $d$ that directly
compares $J$ and $K$, the study-specific treatment effect is expressed
in terms of the direct treatment effect:  
$
\mu_{di}^{JK} = \eta^{JK}_{Dir} + \beta_{di}^{JK};
$
whereas the indirect version of the treatment effect is estimated from
the remaining studies via the consistency equation: 
$
\eta^{JK}_{Ind} = \eta^{AK} - \eta^{AJ}.
$
The posterior distribution of the contrast or inconsistency parameter $\delta^{JK} = \eta^{JK}_{Dir} - \eta^{JK}_{Ind}$ is then examined to check posterior support for the null hypothesis $\delta^{JK} = 0$.

\paragraph{Multiple node-splits}
Although the single node-split approach in \citet{DiasEtAl2010} has been extended to automate the generation of different single node-splitting models for conflict assessment \citep{vanValkenhoefEtAl2016}, the simultaneous splitting of multiple nodes in a NMA has not yet been considered. In section \ref{sec_nmaSplit}, we use multiple splits to investigate conflict in the smoking cessation network beyond heterogeneity, accounting for the multiplicity. 

\subsection{Generalised evidence synthesis \label{sec_ges}}
As further illustration of systematic conflict detection, we consider an evidence synthesis approach to estimating HIV prevalence in Poland, among the exposure group of men who have sex with men (MSM) \citep{RosinskaEtAl2015}. The data aggregated to the national level are given in Supplementary Material Table \ref{tab_plResults}. 
There are three basic parameters to be estimated: the proportion of the male population who are MSM, $\rho$; the prevalence of HIV infection in the MSM group, $\pi$; and the proportion of those infected who are diagnosed, $\kappa$ (Figure \ref{fig_plDAGs}(a)). 

\paragraph{Likelihood}
The total population of Poland, $N = 15,749,944$, is considered
fixed. The remaining 5 data points $y_1,\ldots,y_5$ directly inform,
respectively: $\rho$; prevalence of diagnosed infection $\pi\kappa$;
prevalence of undiagnosed infection $\pi(1-\kappa)$; and lower ($D_L$)
and upper ($D_U$) bounds for the number of diagnosed infections $D =
N\rho\pi\kappa$ (Figure \ref{fig_plDAGs}(a), Supplementary Material Table \ref{tab_plResults}). 
These data are modelled independently as either Binomial ($y_1,y_2,y_3$) or Poisson ($y_4,y_5$).

\paragraph{Priors}
The number diagnosed $D$ is constrained \emph{a priori} to lie between the stochastic bounds $D_L$ and $D_U$, which in turn are given vague log-normal priors. Since $D$ is already defined as a function of the basic parameters, the constraint is implemented via introduction of an auxiliary Bernoulli datum of observed value $1$, with probability parameter given by a functional parameter $c = Pr(D_L \leq D \leq D_U)$ (Figure \ref{fig_plDAGs}(a)). The basic parameters $\rho,\pi$ and $\kappa$ are given independent uniform prior distributions on $[0,1]$. 

\paragraph{Exploratory model criticism}
This initial analysis reveals a lack of fit to some of the data (Supplementary Material Table \ref{tab_plResults}),
with particularly high posterior mean deviances
for the data informing $\rho$ and $\pi\kappa$. This lack of fit in
turn may suggest the existence of conflict in the DAG
\citep{SpiegelhalterEtAl2002}. In \citet{RosinskaEtAl2015}, conflict
between evidence sources was not directly considered or formally
measured, instead resolving the lack of fit by modelling potential
biases in the data in a series of sensitivity analyses. By contrast,
in Section \ref{sec_plSplit} we systematically assess the consistency
of evidence coming from the prior model and from each likelihood
contribution, by splitting the DAG at each functional parameter
(Figure \ref{fig_plDAGs}(b)).

\section{Methods \label{sec_methods}}

\subsection{A single conflict p-value}
Briefly, as in \citet{PresanisEtAl2013}, consider partitioning a DAG $\mathcal{G}(\bs{\phi},\bs{y})$ into two independent
partitions, at a separator node $\phi$. The separator could either be a
founder node, i.e. a basic parameter, or a node internal to the DAG,
and is split into two copies $\phi_a$ and $\phi_b$, one in each
partition (Figure \ref{fig_genSplit}(ii,iii)). Suppose that partition $\mathcal{G}(\bs{\phi}_a, \bs{y}_a)$ contains the data vector $\bs{y}_a$ and provides inference resulting in a posterior distribution $p(\phi_{a} \mid \bs{y}_a)$, and that similarly partition $\mathcal{G}(\bs{\phi}_b, \bs{y}_b)$ results in $p(\phi_{b} \mid \bs{y}_b)$. The aim is to assess the null hypothesis that $\phi_{a} = \phi_{b}$. For $\phi$ taking discrete values, we can directly evaluate $p(\phi_{a} = \phi_{b} \mid \bs{y}_a, \bs{y}_b)$. If the support of $\phi$ is continuous, we consider the posterior probability of $\delta = h(\phi_{a}) - h(\phi_{b})$, where $h(\cdot)$ is a function that transforms $\phi$ to a scale for which a uniform prior is appropriate. The two-sided ``conflict p-value'' is defined as 
$
c = 2 \times \min\left\{\textrm{Pr}\{p_{\delta}(\delta \mid \bs{y}_a, \bs{y}_b) < p_{\delta}(0 \mid \bs{y}_a, \bs{y}_b)\}, 1 - \textrm{Pr}\{p_{\delta}(\delta \mid \bs{y}_a, \bs{y}_b) < p_{\delta}(0 \mid \bs{y}_a, \bs{y}_b)\} \right\}
$,
where $p_{\delta}$ is the posterior density of the difference $\delta$, so that the smaller $c$ is, the greater the conflict.

\subsection{Defining multiple hypothesis tests of conflict}
Generalising now to multiple tests of conflict,
suppose that $\mathcal{G}(\bs{\phi}, \bs{y})$ is partitioned into
$Q$ independent sub-graphs, $\mathcal{G}_1(\bs{\phi}_1, \bs{y}_1),
\ldots, \mathcal{G}_Q(\bs{\phi}_Q, \bs{y}_Q)$, where each disjoint
subset of the data $\bs{y}_q, q \in 1, \ldots, Q$ is chosen to
identify part of the basic parameter space $\bs{\theta}_q =
(\theta_{q1}, \ldots, \theta_{qb_q})$, where $b_q$ is the number
of basic parameters in partition $q$. Note that $\bs{\theta}_q
\subset \bs{\phi}_q$ for each $q \in 1, \ldots, Q$, whereas the
complementary subset $\bs{\phi}_q \setminus \bs{\theta}_q$ consists of
functional and other non-basic parameters. To test the consistency of information provided by each partition about a set of $J$ separator
nodes $(\phi_1^{(s)}, \ldots, \phi_J^{(s)}) \subseteq \bs{\phi}$ from the original model, a set of constrasts 
$
\bs{\delta}_j = (\delta_{j1}, \ldots, \delta_{jC_j})
$
is formed for each $j \in 1, \ldots, J$, one contrast per pair of partitions in which $\phi_j$ appears. A maximum of $Q \choose 2$ contrasts are possible for each separator, i.e. $C_j \leq {Q \choose 2}$. Each contrast $\delta_{jc}$ is defined as
$$
\delta_{jc} = h_{j}(\phi_{jq_A} \mid \bs{y_A}) -
h_{j}(\phi_{jq_B} \mid \bs{y_B})
$$
for the pair of partitions $c = \{q_A,q_B\}$ and node-split
  copies $\{\phi_{jq_A}, \phi_{jq_B}\}$. The functions
$h_{j}(\cdot)$ are functions that transform
the separator nodes $\{\phi_{jq_A}, \phi_{jq_B}\}$ to an appropriate scale for a uniform (Jeffreys’) prior to be applicable, if either is a
founder node in either partition.

Denote the separator nodes in each partition by
$\bs{\phi_q^{(s)}} = \{\phi_{jq}, j \in 1, \ldots, m_q, q \in 1,
  \ldots, Q\}$, where $m_q \leq J$ is the number of separator nodes in
partition $q$. Writing these nodes as a stacked vector
$\bs{\phi_S} = (\bs{\phi_1^{(s)}}, \ldots, \bs{\phi_Q^{(s)}}) =
(\phi_{11}, \ldots, \phi_{m_11}, \phi_{12}, \ldots, \phi_{m_22}, \ldots, \phi_{1Q}, \ldots, \phi_{m_QQ})^T$, and
the transformed version as 
$\bs{\phi_H} = \bs{h}(\bs{\phi_S})$, the total set of contrasts is 
$$
\bs{\Delta} = (\bs{\delta_1}, \ldots, \bs{\delta_J})^T = \bs{C_{\Delta}}^T\bs{\phi_H}
$$
for an appropriate contrast matrix of 1s and 0s,
$\bs{C_{\Delta}}^T$. Note that not every separator node necessarily
appears in every partition, so although $\bs{\phi_H}$ has maximum
length $J \times Q$, in practice, its length 
$m = \sum_{q=1}^Q m_q \leq J \times Q$. The contrast matrix $\bs{C_{\Delta}}^T$ therefore
has dimension $p \times m$, so that it maps from the space of the $m$
separator nodes (including node-split copies) to that of the
$p = \sum_{j=1}^J C_j$ contrasts. A test for consistency of the information in each partition may be expressed as a test of the null hypothesis that
\begin{equation}
H_0: \bs{\Delta} = \bs{C_{\Delta}}^T\bs{\phi_H} = \bs{0} \label{eqn_nullhyp_bayes}
\end{equation}

\subsection{Asymptotic theory}

Using standard asymptotic theory \citep[][see also derivation in Supplementary Material Appendix \ref{sec_AppBasymp}]{BernardoSmith1994}, it can be
shown that if the joint posterior distribution of all parameters $\bs{\phi}$
  in all partitions is asymptotically multivariate normal (i.e. if the
  prior is flat enough relative to the likelihood),
and if
$\frac{\partial\bs{\Delta(\phi)}}{\partial\bs{\phi}} = \bs{C_{\Delta}}^T$ is
non-singular with continuous entries, then the posterior mean of
$\bs{\Delta}$ is $\overline{\bs{\Delta}} = \bs{C_{\Delta}}^T
\overline{\bs{\phi_H}} \overset{a}{\approx} \bs{C_{\Delta}}^T
\hat{\bs{\phi}}_H$ and the posterior variance-covariance matrix of
$\bs{\Delta}$ is $\bs{S_{\Delta}} \overset{a}{\approx}
\bs{C_{\Delta}}^T \bs{V_H} \bs{C_{\Delta}}$, where: $\bs{\hat{\phi}}_H$ is the
maximum likelihood estimate of $\bs{\hat{\phi}}_H$; the matrix $\bs{V_H} =
  \bs{J_h}(\bs{\hat{\phi}_S})^T \bs{V_S} \bs{J_h}(\bs{\hat{\phi}_S})$;
  $\bs{J_h}(\bs{\hat{\phi}_S})$ is the Jacobian of the transformation
  $\bs{h}(\bs{\phi_S})$; and $\bs{V_S}$ is a blocked
diagonal matrix consisting of the inverse observed information
matrices for the separator nodes in each partition along the diagonal. The posterior summaries $\overline{\bs{\Delta}}$ and $\bs{S_{\Delta}}$, i.e. the Bayes' estimator under a mean-squared error Bayes' risk function and corresponding variance-covariance matrix, may therefore be used under the general simultaneous inference framework of \citet{HothornEtAl2008,BretzEtAl2011} to construct a multiplicity-adjusted test that $\bs{\Delta} = \bs{0}$. 

\subsection{Simultaneous hypothesis testing}

Given the estimator $\overline{\bs{\Delta}}$ and corresponding variance-covariance matrix $\bs{S_{\Delta}}$, define a vector of test statistics $\bs{T}_n = \bs{D}_n^{-1/2} (\overline{\bs{\Delta}} - \bs{\Delta})$,
where $n$ is the dimension of the data $\bs{y}$ and $\bs{D}_n = diag(\bs{S_{\Delta}})$. Then it can be shown \citep{HothornEtAl2008,BretzEtAl2011} that $T_n$ tends in distribution to a multivariate normal distribution,
$
\bs{T}_n \asym N_m(\bs{0}, \bs{R}) \label{eqn_T_n_refdist}
$, where $\bs{R} := \bs{D}_n^{-1/2} \bs{S_{\Delta}} \bs{D}_n^{-1/2} \in \mathbb{R}^{m,m}$ is the posterior correlation matrix for the vector (length $m$) of contrasts $\bs{\Delta}$. Under the null hypothesis (\ref{eqn_nullhyp_bayes}), $\bs{T}_n = \bs{D}_n^{-1/2} \overline{\bs{\Delta}} \asym N_m(\bs{0}, \bs{R})$,
and hence, assuming $\bs{S_{\Delta}}$ is fixed and known, the authors show that a global $\chi^2$-test of conflict can be formulated:
$$
X^2 = \bs{T}_n^T\bs{R}^+\bs{T}_n \tendist \chi^2(Rank(\bs{R}))
$$
where the superscript $^+$ denotes the Moore-Penrose inverse of the corresponding matrix and $Rank(\bs{R})$ is the degrees of freedom. Importantly, it is also possible to construct multiply-adjusted local (individual) conflict tests, based on the $m$ $z-$scores corresponding to $\bs{T}_n$ and the null distribution of the maximum of these, $Z_{max}$, \citep{HothornEtAl2008,BretzEtAl2011}. This latter null distribution is obtained by integrating the limiting $m-$dimensional multivariate normal distribution over $[-z,z]$ to obtain the cumulative distribution function $\mathbb{P}(Z_{max} \leq z)$. The individual conflict p-values are then calculated as $\mathbb{P}(|z_k| < Z_{max}), k \in 1, \ldots, m$, with a corresponding global conflict p-value (an alternative to the $\chi^2$-test) given by $\mathbb{P}(|z_{max}| < Z_{max})$.

\section{Examples \label{sec_applications}}

We now illustrate the idea of systematic multiple node-splitting to assess conflict in our two motivating examples. All analyses were carried out in \texttt{OpenBUGS 3.2.2} \citep{LunnEtAl2009} and \texttt{R 3.2.3} \citep{Rproject2015}. We use the \texttt{R2OpenBUGS} package \citep{R2OpenBUGS2005} to run \texttt{OpenBUGS} from within \texttt{R} and the \texttt{multcomp} package \citep{BretzEtAl2011} to carry out the simultaneous local and global max-T tests.

\subsection{Network meta-analysis}

\begin{figure}
  \centering
  \includegraphics[width = 0.85\textwidth]{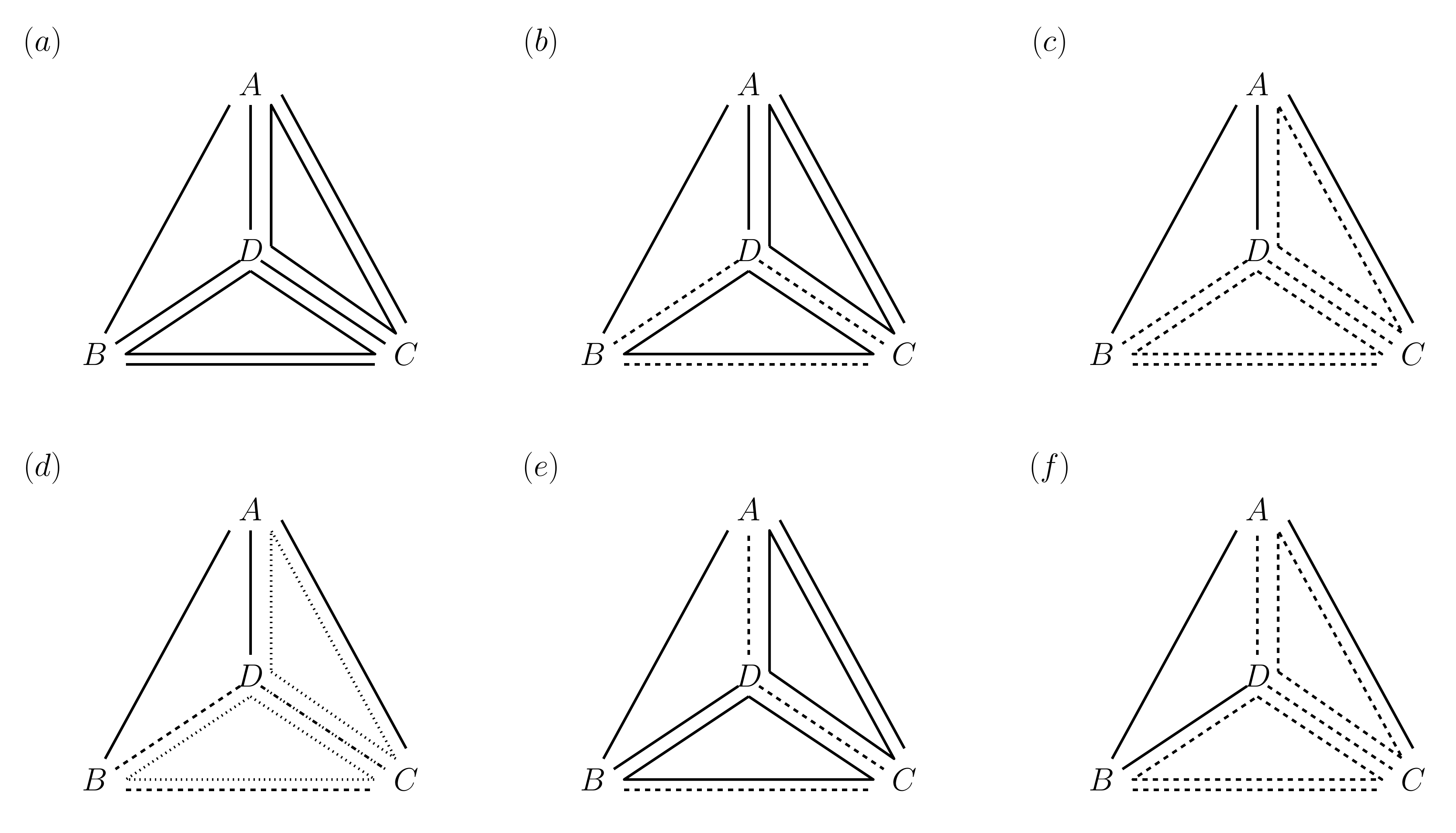}
  \caption{Smoking cessation evidence network, under (a) a consistency assumption; (b)-(f) inconsistency assumptions, where the evidence is partitioned in different ways. In (b), (c), (e) and (f), the direct evidence (dashed lines) is compared with the indirect evidence (solid lines) on each contrast where there is a dashed line. In (d), the evidence is separated into three spanning trees and a fourth partition for the multi-arm trial evidence.
\label{fig_smkSpan}}
\end{figure}

Consider first a NMA in general, and for simplicity, assume 
there are no multi-arm trials and a common-effect model  (equation
(\ref{eqn_consFE})) for the data. The basic parameters $\bs{\eta}_b$
form a spanning tree of the network of evidence (Figure
\ref{fig_smkSpan}), i.e. a graph with no cycles, such that each node
in the network can be reached from every other node, either directly
or indirectly through other nodes
\citep{vanValkenhoefEtAl2012}. Multiple possible partitionings of the
evidence network exist, so a choice must be made (Figure
\ref{fig_smkSpan}). 
Suppose the spanning tree $\bs{\eta}_b$ is identifiable by a set of evidence $\bs{Y}_b$ containing outcomes from all trials designed to directly estimate the treatment effects in $\bs{\eta}_b$. Then every treatment effect is identifiable from $\bs{Y}_b$, by definition of a spanning tree and the fact that each treatment effect represented by edges outside the spanning tree is a functional parameter in the set $\bs{\eta}_f$, equal to a linear combination of the basic parameters. The data $\bs{Y}_b$ therefore \emph{indirectly} inform the functional parameters $\bs{\eta}_f$, whereas the remaining data, $\bs{Y}_f = \bs{Y} \setminus \bs{Y}_b$ \emph{directly} inform $\bs{\eta}_f$. A comparison between the direct and indirect evidence on $\bs{\eta}_f$ is therefore possible, to assess conflict between the two types of evidence. The network is split into two partitions, $\{\bs{\eta}_f^{Dir},\bs{Y}_f\}$ (the ``direct evidence partition'', DE) and $\{\bs{\eta}_f^{Ind},\bs{Y}_b\}$ (the ``spanning tree partition'', ST) and the direct and indirect versions of the functional parameters compared:
$
\bs{\Delta} = \bs{\eta}_f^{Dir} - \bs{\eta}_f^{Ind} .
$
Depending on the studies that are in the DE partition, the
basic parameters $\bs{\eta}_b$ may also be weakly identifiable in
the DE partition, due to prior information. Since a NMA model may be
formulated as a DAG, this Direct/Indirect partitioning is equivalent
to a multi-node split in the DAG at the functional parameters (Supplementary Material Figure \ref{fig_nmaDAGsplit}).

Generalising now to more complex situations, if the direct data $\bs{Y}_f$ form a sub-network of evidence, the question arises of whether these data should be split into further partitions, by identifying a spanning tree for the sub-network. Then the vector $\bs{\Delta}$ of contrasts to test would involve comparisons between more than two partitions, e.g. for three partitions:
$$
\bs{\Delta} = \left( \bs{\eta}_f^1 - \bs{\eta}_f^2, \bs{\eta}_f^1 - \bs{\eta}_f^3, \bs{\eta}_f^2 - \bs{\eta}_f^3 \right)^T
$$

If we now consider a random rather than common heterogeneity effects model (equation (\ref{eqn_consRE})), a decision must be made on how to handle the variance components in $\Sigma_{\beta}$. One approach would be to split the variance components simultaneously with the means, so that $\bs{\Delta}$ also includes contrasts for the variances. Alternatively, if the variance components are not well identified by the evidence in a partition, a common variance component could be assumed. Such commonality could potentially allow for feedback between partitions, since they would not be fully independent \citep{MarshallSpiegelhalter2007,PresanisEtAl2013}.

Finally, for multi-arm trials, the key consideration is that multi-arm studies should have internal consistency, and hence their observations should not be split between partitions. A choice must therefore be made whether to initially include multi-arm data in the ST data $\bs{Y}_b$, in the DE data $\bs{Y}_f$, or in a third partition of their own. In the latter case, any study-specific treatment effect $\mu^{JK}_{di}$, where $d$ is a multi-arm design, could be compared at least with the ST partition, where $\eta^{JK}$ is definitely identified. Potentially, it could also be compared simultaneously with the DE partition, if the edge $JK$ is identifiable in the DE partition. The comparison can be made even if $JK$ is not identifiable, or only weakly identifiable from the prior, but if the prior is diffuse, then no conflict will be detected due to the uncertainty. Such a comparison is not therefore particularly meaningful, unless we are interested in prior-data conflict.

\paragraph{Smoking cessation example \label{sec_nmaSplit}}
To illustrate concretely the above issues, we consider first the spanning tree $(AB,AC,AD)$ corresponding to the parameters $\bs{\eta_b} = \{\eta^{AB},\eta^{AC},\eta^{AD}\}$ for the smoking cessation example. Figures \ref{fig_smkSpan}(b-d) demonstrate different ways of splitting the evidence based on this spanning tree, depending on how we treat the evidence from multi-arm trials. In Figures \ref{fig_smkSpan}(b,c), we consider just two partitions, with the multi-arm evidence either left in the ST partition $\{\bs{\eta}_f^{Ind},\bs{Y}_b\}$ or included in the DE partition $\{\bs{\eta}_f^{Dir},\bs{Y}_f\}$, respectively. We compare the direct and indirect evidence on each of the edges or treatment comparisons $(BC,BD,CD)$. In Figure \ref{fig_smkSpan}(d), we consider a series of spanning trees ($(AB,AC,AD), (BC,BD)$ and $(CD)$), together with a final partition consisting of evidence from multi-arm trials, resulting in four partitions.

We also consider an alternative choice of spanning tree, $(AB,AC,BD)$, as in Figures \ref{fig_smkSpan}(e,f). In these two models, we again make a choice between including the multi-arm evidence in either the ST or DE partitions and compare the evidence in each partition on edges $(AD,BC,CD)$. In all cases, we assume random heterogeneity effects and make the choice to assume common variance components across the partitions, splitting only the means.

Table \ref{tab_nmaResults} gives posterior mean (sd) estimates of the treatment effects (log odds ratios) for edges outside the spanning tree, from each partition, where the subscript 1 denotes the ST partition and 2 denotes the DE partition for the two-partition models (b,c,e,f). For the four-partition model (d), 1-3 denote the sequential spanning tree partitions and 4 the multi-arm trial partition. Also given, for each edge outside the original spanning tree, are the posterior mean (sd) differences between partitions and both the local and global posterior probabilities of no difference, adjusted for the multiple tests and their correlation. First, note that the global test of no conflict varies by model, and hence by what partitions of evidence are compared with each other: the posterior probability of no conflict in model (b) is $94.7\%$, compared to only $23.4\%$ and $27.4\%$ for models (c) and (e). These latter two models appear to detect some mild evidence of conflict, despite the large uncertainty in many of the partition-specific treatment effect estimates, with several of the posterior standard deviations of the same order of magnitude as the corresponding posterior means, if not larger. The DIC is also slightly smaller for the two models (c) and (e) which detect potential conflict, compared to those that don't. This lack of invariance of the global test to the partitions employed suggests it is not enough to rely on a single node-splitting model to search for conflict in a DAG. Moreover, it motivates looking at local tests for conflict in different node-splitting models, to locate the specific items of evidence that may conflict with each other.

A closer look at the local posterior probabilities of no conflict for
each edge outside the initial spanning tree reveals that the potential
conflict detected by models (c) and (e) involves edges including
treatment $D$ (posterior probabilities $17.8\%$ and $18.6\%$ for edges
$BD$ and $CD$ in model (c), $12.4\%$ and $10.5\%$ for
edges $AD$ and $CD$ in model (e)). Each of these four local tests
involves a partition where the estimated treatment effect for the
relevant edge is implausibly large ($>6$ on the log odds ratio scale,
i.e. $>400$ on the odds ratio scale) %. In model (e), the estimate in partition 2 for the $AD$ edge of $12.7 (6.2)$ is only directly informed by a single study, whose sample sizes are among the smallest in this network (study 7 in Supplementary Material Table 1), with a $0$ numerator for treatment $A$. In general, the studies included in partition 2 of model (e) (studies 7, 20, 23 and 24 in Supplementary Material Table 1) are amongst the smallest sample sizes of the network. 
and where the sample sizes of the studies involved are small (e.g. studies 7, 20, 23 and 24 in Supplementary Material Table \ref{tab_smkData}).
%Note that by including the multi-arm studies in the spanning tree partition (Figure \ref{fig_smkSpan}(e)), model (e) is equivalent to a model with a different spanning tree, $(AD,CB,CD)$, where the multi-arm trials are included in the partition outside of the spanning tree. These results suggest that those studies contributing most to the detected conflict are those with small sample sizes, leading to implausible and uncertain estimates.

Unlike models (c), (e) and (f), where in both partitions, each sub-network spans all 4 treatments, in models (b) and (d), the spanning tree chosen, $(AB,AC,AD)$, is such that for each sub-network outside the spanning tree, not all the treatments are included (Figure \ref{fig_smkSpan}). This results in a lack of identifiability for the basic parameters $\bs{\eta_b}$ in partition 2 of model (b) and in partitions 2 and 3 of model (d) (Table \ref{tab_nmaResults}), where their estimates are dominated by their diffuse prior distribution (Normal$(0,10^2)$ on the log odds ratio scale). There is therefore no potential for detecting conflict about the basic parameters $\bs{\eta_b}$, only about the functional parameters $\bs{\eta_f}$.

The different results obtained from each of the five models are understandable, since each model partitions the evidence in a different way, and the detection of conflict relies on the conflicting evidence being in different rather than the same partitions. However, where the same evidence is in the same partition for different models --- for example, the evidence directly informing the $AC$ edge in models (c) and (d) --- approximately the same estimate is reached in each model, as expected ($0.81 (0.26)$ in model (c), $0.82 (0.28)$ in model (d), Table \ref{tab_nmaResults}).

\subsection{HIV prevalence evidence synthesis \label{sec_plSplit}}

\begin{figure}
  \centering
  \includegraphics[width = 0.85\textwidth]{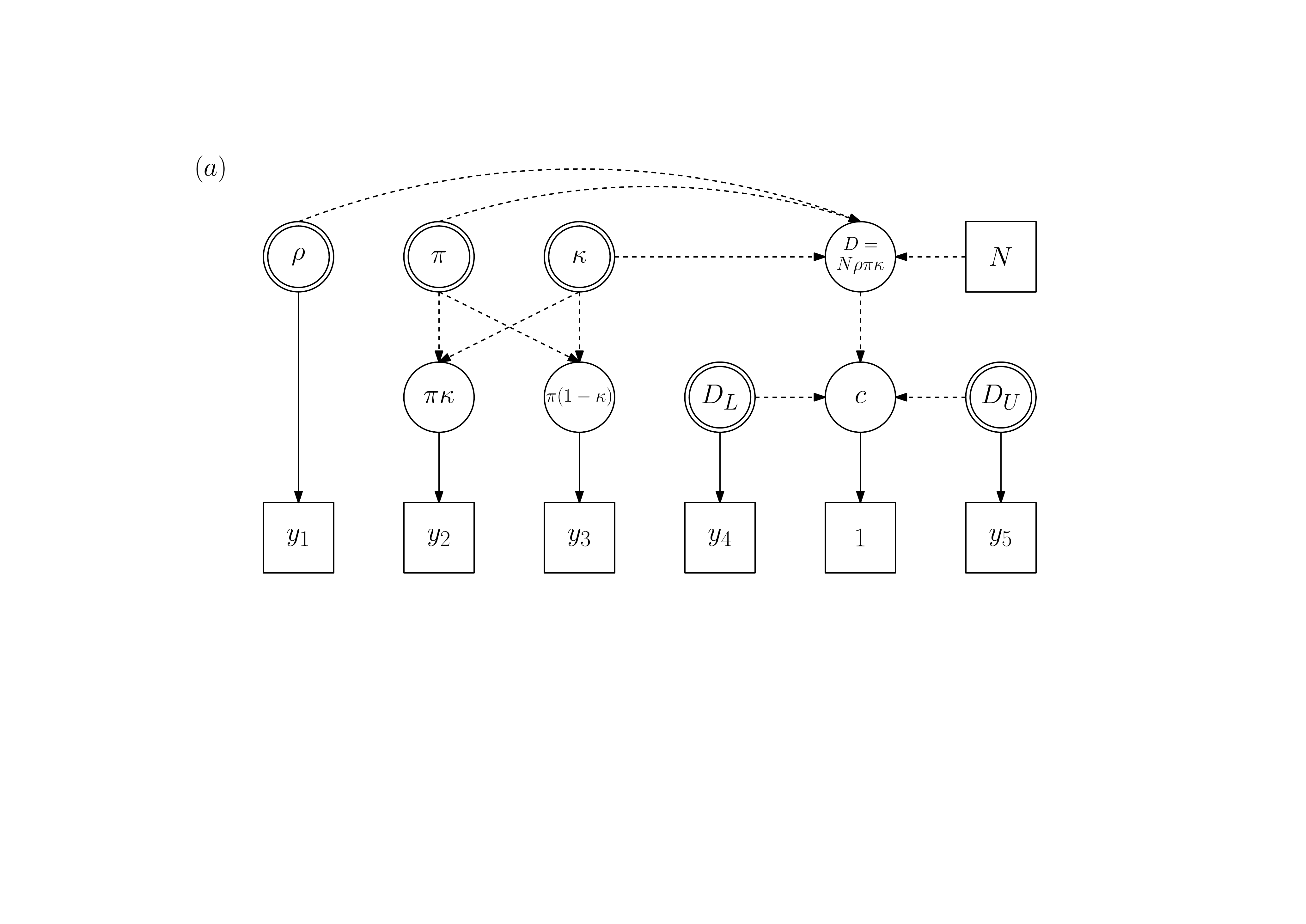}
  \includegraphics[width = 0.85\textwidth]{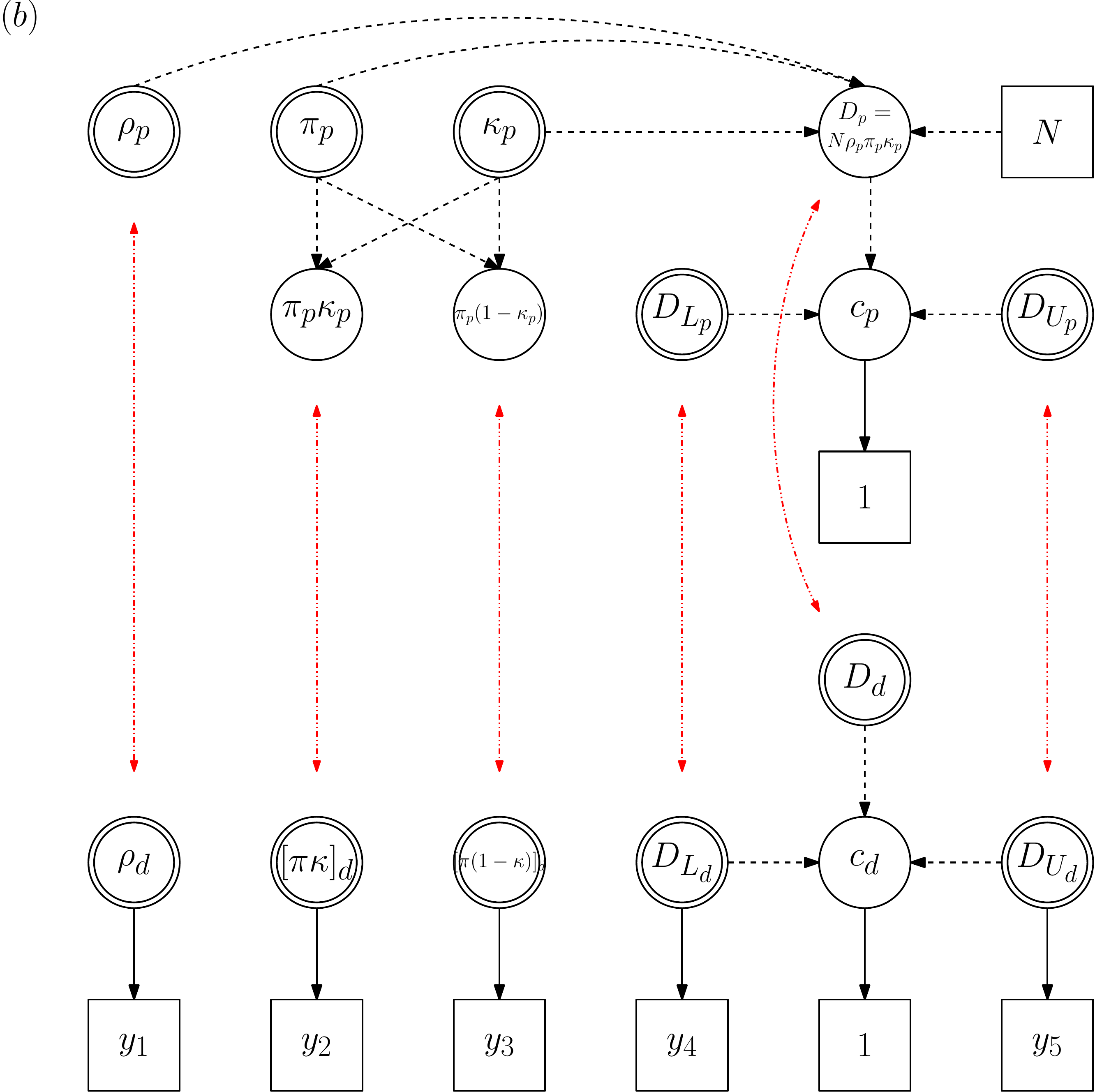}
  \caption{(a) DAG of initial model for synthesising Polish HIV
    prevalence data. (b) DAG of multiple node-split model comparing
    priors to each likelihood contribution. Note that the square
    brackets are used in denoting the nodes in the likelihood
    partition ($[\pi\delta]_d, [\pi(1-\delta)]_d$) to emphasise the
    fact that these two nodes are independent parameters not
    functionally related to each other. \label{fig_plDAGs}}
\end{figure}

Figure \ref{fig_plDAGs}(b) demonstrates the multiple node-splits we make to systematically assess conflict in the original DAG of Figure \ref{fig_plDAGs}(a), separating out the contributions of the prior model and each likelihood contribution. These node-splits result in 5 partitions, with 6 contrasts to test for equality to zero. Denoting the nodes in the ``prior'' partition (above the red arrows in Figure \ref{fig_plDAGs}(b)) by the subscript $p$ and the nodes in each ``likelihood'' partition (below the red arrows in Figure \ref{fig_plDAGs}(b)) by $d$, the vector of contrasts to test is then
\begin{eqnarray*}
\bs{\Delta} & = & (h(\rho_p) - h(\rho_d), h(\pi_p\kappa_p) - h([\pi\kappa]_d), h(\pi_p(1-\kappa_p)) - h([\pi(1-\kappa)]_d), \\
& & g(D_{L_p}) - g(D_{L_d}), g(D_{U_p}) - g(D_{U_d}), g(D_p) - g(D_d)))^T
\end{eqnarray*}
where $h(\cdot)$ and $g(\cdot)$ denote the logit and log functions respectively. These contrasts are represented by the red dot-dashed arrows in Figure \ref{fig_plDAGs}(b). In the ``prior'' partition, the priors given to the basic parameters
are those of the original model (Section \ref{sec_ges}).
In each ``likelihood'' partition, the basic parameters are given
Jeffreys' priors so that the posteriors represent only the likelihood. These priors are Beta$(\nicefrac{1}{2}, \nicefrac{1}{2})$ for the proportions and $p(D_{B_d}) \propto 1 / D_{B_d}^{1/2}$ for the lower and upper bounds ($B = L,U$) for $D$. $D_d$ is given a Uniform prior between $D_{L_d}$ and $D_{U_d}$.

\begin{figure}
  \centering
  \includegraphics[width = \textwidth]{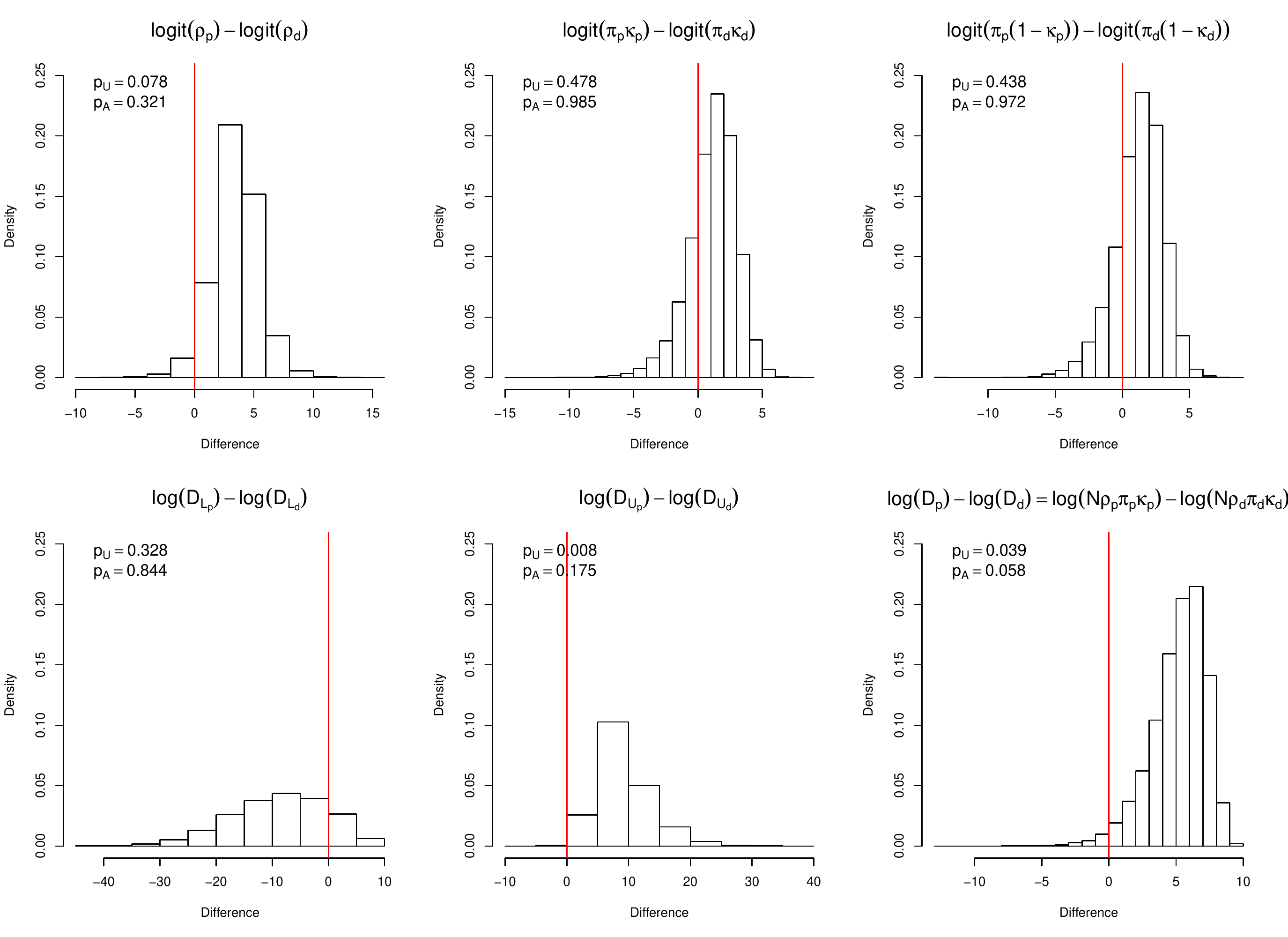}
  \caption{Posterior distributions of the contrasts $\bs{\Delta}$ for the HIV prevalence example. The red lines denote 0 difference, $p_U$ is the unadjusted and $p_A$ the multiply-adjusted individual conflict p-value respectively. \label{fig_plSatDiffs}}
\end{figure}

Figure \ref{fig_plSatDiffs} shows the posterior distributions of the
contrasts $\bs{\Delta}$, where 0 lies in these distributions and the
corresponding unadjusted ($p_U$) and multiply-adjusted ($p_A$)
individual conflict p-values testing for equality to $0$. A global
$\chi^2$-squared (Wald) test gives a conflict p-value of $0.001$,
suggesting conflict exists somewhere in the DAG. Examining the
individual unadjusted (naive) conflict p-values would suggest
prior-data conflict at the upper bound for the number diagnosed $D_U$
(posterior probability of zero difference is $p_U = 0.008$) and
hence at the number diagnosed itself, $D$ ($p_U = 0.039$), as well as
possibly at the proportion at risk, $\rho$ ($p_U = 0.078$). However,
once the correlation between the individual tests has been taken into
account, the posterior probabilities of no conflict increase for all contrasts, albeit the
probabilities are still low for $D_U$ and $D$, at $p_A = 0.175$ and
$p_A = 0.058$ respectively. Note that the posterior
contrasts in Figure \ref{fig_plSatDiffs} are slightly non-normal, hence
we interpret the adjusted posterior probabilities of no conflict as
exploratory, rather than as absolute measures. 

Examining closer the posterior distributions of the ``prior'' and
``likelihood'' versions of the node $D$ 
(Supplementary Material Figure \ref{fig_plSatD}, upper panel), we
visualise better the prior-data conflict: the ``likelihood'' version
lies very much in the lower tail of the ``prior'' version. This is in
spite of -- or rather because of -- the flat Uniform priors of the
prior model, which translate into a non-Uniform implied prior for the
function $D_p = N\rho_p\pi_p\kappa_p$.

The ``saturated'' model splitting apart each component of evidence in
the DAG allows us to assess prior-data conflict in this model, but not
conflict between different combinations of the likelihood evidence,
due to lack of identifiability: in each likelihood partition in Figure
\ref{fig_plDAGs}(b), clearly only the parameter directly informed by
the data, whether basic or functional, can be identified. To assess
consistency of evidence between likelihood terms, we employ a
cross-validatory ``leave-n-out'' approach, for $n = 1$ and $n = 2$,
splitting in each case the relevant nodes \emph{directly} informed by
the left-out data items. Note that other possibilities exist, such as splitting
at the basic parameters, depending on which data are left
out. Table \ref{tab_plResultsLnO} gives unadjusted ($p_U$) and various
multiply-adjusted ($p_{AW},p_{AL},p_{AA}$) individual posterior
probabilities of no difference between nodes split between partitions
1 (the ``left-out'' evidence) and 2 (the remaining evidence). These posterior probabilities highlight inconsistency in the network of evidence $\{y_1,y_2,y_4,y_5\}$, i.e. informing the three nodes $\rho, \pi\kappa$ and $D = N\rho\pi\kappa$. Splits at these three nodes demonstrate low posterior probabilities of no difference in the ``leave-1-out'' models (A), (B) and (E), and in the ``leave-2-out'' models (B), (C), and (J) in particular. There is no potential for the evidence $y_3$ on the prevalence of undiagnosed infection $\pi(1-\kappa)$ to conflict with any other evidence, since $\pi$ and $\kappa$ are not separately identifiable from the remaining evidence $\{y_1,y_2,y_4,y_5\}$ alone. Hence all of the posterior probabilities of no difference concerning the node $\pi(1-\kappa)$ are high.

The conflict in the $\{y_1,y_2,y_4,y_5\}$ network is well illustrated
by the node-split model (J), where the count data on the lower and
upper bounds for the $D$ are ``left out'' in partition 1.
Supplementary Material Figure \ref{fig_plSatD} (lower panel) shows the
posterior distributions for each of $D_L, D_U$ and $D$ in both
partitions. 
Since in partition 2 the data on
the limits for $D$ have been excluded, the posterior distributions for
the bounds (solid black and red lines) are flat and hugely variable. Despite this, the posterior
distribution for $D_2$ is relatively tightly peaked, due to the
indirect evidence on $D_2$ provided by the data informing $\rho_2$ and
$\pi_2\kappa_2$. It is this indirect evidence that conflicts with the
direct evidence informing $D_1$ via the data $\{y_4,y_5\}$ on the
bounds for $D_1$. 

\section{Discussion \label{sec_discuss}}

We have proposed here the systematic assessment of conflict in an
evidence synthesis, in particular accounting for the multiple tests
for consistency entailed, through the simultaneous inference framework
proposed by \citet{HothornEtAl2008,BretzEtAl2011}. We have chosen the max-T tests that allow both for multiply-adjusted local and global testing simultaneously. 

Note that the use of this (typically classical) simultaneous inference
framework relies on the asymptotic multivariate normality of the joint
posterior distribution. In cases where the likelihood does not
dominate the prior, resulting in a skewed or otherwise non-normal
posterior, we treat the results of conflict analysis as exploratory,
rather than absolute measures of conflict. If
the posterior is skewed but still uni-modal, a global, implicitly
multiply-adjusted, test for conflict can be formulated in terms of the
Mahalanobis distance of each posterior sample from their mean, as we
proposed in \citet{PresanisEtAl2013}. This is a multivariate
equivalent of calculating the tail area probability for regions
further away from the posterior mean than the point $\bs{0}$. However,
the Mahalanobis-based test does not allow us to obtain local tests for
conflict, nor does it apply in the case of a multi-modal posterior. In
the latter case, kernel density estimation could be used to
obtain the multivariate tail area probability, although such
estimation is computationally challenging for large posterior dimension.

Although generalised evidence syntheses have mostly been carried out in a Bayesian framework, there are examples \citep[e.g.][]{CommengesHejblum2013} that are either frequentist or not fully Bayesian. In the NMA field, maximum likelihood and Bayesian methods are both common \citep[e.g.][]{WhiteEtAl2012,JacksonEtAl2014}. An advantage of the simultaneous inference framework \citep{HothornEtAl2008,BretzEtAl2011} is that, given any estimator $\bs{\overline{\Delta}}$ of a vector of differences and its corresponding variance-covariance matrix $\bs{S_{\Delta}}$, regardless of the method used to obtain the estimates, the global and local max-T tests can be formulated. 

Conflict p-values can be seen as
cross-validatory posterior predictive checks
\citep{PresanisEtAl2013}. There is a large literature on various types
of Bayesian predictive diagnostics, including prior-, posterior- and
mixed-predictive checks
\citep[e.g.][]{Box1980,GelmanEtAl1996,MarshallSpiegelhalter2007}. A
key issue much discussed in this literature is the lack of uniformity
of posterior predictive p-values under the null hypothesis
\citep{Gelman2013}, with such p-values conservative due to the double
use of data. Much work has therefore been devoted to either
alternative p-values \citep[e.g.][]{BayarriBerger2000} or
post-processing of p-values to calibrate them
\citep[e.g.][]{SteinbakkStorvik2009}. \citet{Gelman2013} argues that
the importance of uniformity depends on the context in
which the model checks are conducted: in general non-uniformity is
not an issue, but if the posterior predictive tests rely on
parameters or imputed latent data, then care should be taken. Since conflict p-values are cross-validatory, the issue of conservatism and the double use of data does not apply. In fact, for a wide class of standard hierarchical models, \citet{Gasemyr2015} has demonstrated the uniformity of the conflict p-value. 

As illustrated by both applications, the choice of different ways of
partitioning the evidence in a DAG can lead to different conclusions
over the existence of conflict. This is to be expected when
considering the local conflict p-values, since conflicting evidence
may need to be in different partitions in order to be detectable. This
is analogous to the idea of ``masking'' in cross-validatory outlier
detection, where outliers may not be detected if multiple outliers
exist \citep{ChalonerBrant1988}. In the case of the global tests for
conflict, the NMA example showed that these are also not invariant to
the choice of partition. In the NMA literature, alternative methods
accounting for inconsistency include models that introduce
``inconsistency parameters'' that absorb any variability due to
conflict beyond between-study heterogeneity
\citep{LuAdes2006,HigginsEtAl2012,JacksonEtAl2014}. \citet{HigginsEtAl2012,JacksonEtAl2014}
have pointed out that the apparent algorithm that \citet{LuAdes2006}
follow for identifying inconsistency parameters does not guarantee
that all such parameters are identified, nor that the Lu-Ades model is
invariant to the choice of baseline treatment. The authors further
posit, and more recently have proved \citep{JacksonEtAl2015}, that
their ``design-by-treatment interaction model'', which introduces an
inconsistency parameter systematically for each non-baseline treatment
within each design, contains each possible Lu-Ades model as a
sub-model. In related ongoing work, we note that each Lu-Ades model corresponds to a particular choice of node-splitting model, one being a reparameterisation of the other. The lack of invariance of results of testing for inconsistency from one Lu-Ades model to another is therefore not surprising, since, as we illustrated here, different choices of node-splitting model correspond to different partitions of evidence being compared. The lack of invariance of a global test for conflict to the choice of node-splitting model, although unsurprising, is perhaps unsatisfactory: however, as we illustrated in this paper, this lack clearly emphasises the need for a more comprehensive and systematic assessment of conflict throughout a DAG, both at a local level and across different types of node-split model, than just a single global test can provide. We therefore recommend that although a global test may be an initial step in any conflict analysis, to be sure of detecting any potential conflict requires testing for conflict throughout a DAG. One strategy is to start from splitting every possible node in the DAG, as we did in the HIV example, before looking at more targeted leave-n-out approaches. The design-by-treatment interaction model provides a way of doing so and we are further investigating the relationship of the (fixed inconsistency effects) design-by-treatment interaction model to such a ``saturated'' node-splitting model. 

Note that in the NMA example considered here, we have concentrated on
a ``contrast-based'' as opposed to ``arm-based'' parameterisation
\citep{HongEtAl2015,DiasAdes2015}. Also, we have considered
the case where each study has a study-specific baseline treatment
$B_d$ and the network as a whole has a baseline treatment
$A$. However, alternative parameterisations could be considered, such
as using a two-way linear predictor with main effects for both
treatment and study, treating the counter-factual or missing treatment
designs as missing data \citep{JonesEtAl2011,PiephoEtAl2012}. Although
we have not yet explored alternative parameterisations, we posit that
systematic node-splitting could be equally well applied.

As with any cross-validatory work, the systematic assessment of conflict at every node in a DAG can quickly become computationally burdensome as a model grows in dimension. An area for future research is the systematic analysis of conflict using efficient algorithms \citep{LunnEtAl2013,GoudieEtAl2015} in a Markov melding framework \citep{GoudieEtAl2016} which allows for an efficient modular approach to model building.

\section*{Acknowledgements}
This work was supported by the Medical Research Council [Unit
Programme number U105260566]; and the Polish National Science Centre
[grant no. DEC-2012/05/E/ST1/02218]. The authors also thank Ian White
and Dan Jackson for their very helpful comments.

 \bibliographystyle{chicago} 
 \bibliography{refs}

%\clearpage

\begin{table}[!p]
  \caption{Multiply adjusted posterior mean (sd) estimates of conflict between partitions, for each model (b)-(f) respectively. In the two-partition models (b,c,e,f), partition 1 is the spanning tree (indirect) evidence partition and partition 2 is the direct data partition. In model (d), partitions 1-3 are the sequential spanning trees and partition 4 is the multi-arm study partition. \label{tab_nmaResults}}
  \centering
  \scriptsize
  \begin{tabular}{lrrrrrrrrrr}
    \toprule
    \midrule
    ST: & \multicolumn{6}{c}{\tiny{AB,AC,AD}} & \multicolumn{4}{c}{\tiny{AB,AC,BD}} \\
    Model: & \multicolumn{2}{c}{(b)} & \multicolumn{2}{c}{(c)} & \multicolumn{2}{c}{(d)} & \multicolumn{2}{c}{(e)} & \multicolumn{2}{c}{(f)} \\
    Posterior: & Mean & SD  & Mean & SD & Mean & SD & Mean & SD & Mean & SD  \\
    \midrule
           $AB_1$ & 0.472 & (0.489) & 0.338 & (0.534) & 0.329 & ( 0.568) &  0.259 & (0.429) & 0.334 & (0.566) \\
           $AB_2$ &-0.415 & (5.276) & 0.319 & (0.983) &-0.230 & ( 5.849) &  6.261 & (3.251) & 1.513 & (1.041) \\
           $AB_3$ &       &         &       &         &-0.044 & (10.009) &        &         &       &         \\
           $AB_4$ &       &         &       &         & 0.456 & ( 1.247) &        &         &       &         \\
    \midrule
           $AC_1$ & 0.877 & (0.262) & 0.814 & (0.261) & 0.828 & ( 0.280) &  0.812 & (0.238) & 0.824 & (0.272) \\
           $AC_2$ &-0.165 & (5.262) & 0.615 & (0.866) &-0.379 & ( 5.848) &  6.173 & (3.140) & 1.496 & (0.835) \\
           $AC_3$ &       &         &       &         & 0.114 & ( 7.051) &        &         &       &         \\
           $AC_4$ &       &         &       &         & 0.784 & ( 0.956) &        &         &       &         \\
     \midrule
           $AD_1$ & 1.010 & (0.598) & 9.337 & (4.999) & 9.508 & ( 5.330) &  0.908 & (0.794) & 1.748 & (1.526) \\
           $AD_2$ & 0.262 & (5.266) & 0.679 & (0.871) & 0.871 & ( 5.859) & 12.712 & (6.225) & 3.102 & (1.690) \\
           $AD_3$ &       &         &       &         & 0.319 & ( 7.044) &        &         &       &         \\
           $AD_4$ &       &         &       &         & 0.439 & ( 0.956) &        &         &       &         \\
$\Delta_{AD_{1-2}}$ &       &         &       &         &       &          &-11.804 & (6.268) &-1.354 & (2.279) \\
     $p_{AD_{1-2}}$ &       &         &       &         &       &          &  0.124 &         & 0.806 &         \\
    \midrule
           $BC_1$ & 0.405 & (0.527) & 0.476 & (0.590) & 0.499 & ( 0.633) &  0.553 & (0.461) & 0.490 & (0.629) \\
           $BC_2$ & 0.251 & (0.808) & 0.296 & (0.577) &-0.149 & ( 1.015) & -0.087 & (0.951) &-0.017 & (0.693) \\
           $BC_3$ &       &         &       &         & 0.158 & (12.270) &        &         &       &         \\
           $BC_4$ &       &         &       &         & 0.329 & ( 0.957) &        &         &       &         \\
$\Delta_{BC_{1-2}}$ & 0.155 & (0.963) & 0.180 & (0.821) & 0.649 & ( 1.198) &  0.641 & (1.059) & 0.507 & (0.937) \\
$\Delta_{BC_{1-3}}$ &       &         &       &         & 0.341 & (12.290) &        &         &       &         \\
$\Delta_{BC_{1-4}}$ &       &         &       &         & 0.171 & ( 1.161) &        &         &       &         \\
     $p_{BC_{1-2}}$ & 0.986 &         & 0.971 &         & 0.979 &          &  0.807 &         & 0.839 &         \\
     $p_{BC_{1-3}}$ &       &         &       &         & 1.000 &          &        &         &       &         \\
     $p_{BC_{1-4}}$ &       &         &       &         & 1.000 &          &        &         &       &         \\
    \midrule
           $BD_1$ & 0.538 & (0.691) & 8.999 & (5.031) & 9.180 & ( 5.357) &  0.649 & (0.530) & 1.414 & (1.173) \\
           $BD_2$ & 0.678 & (0.809) & 0.360 & (0.569) & 1.101 & ( 1.026) &  6.451 & (3.083) & 1.589 & (0.802) \\
           $BD_3$ &       &         &       &         & 0.363 & (12.255) &        &         &       &         \\
           $BD_4$ &       &         &       &         &-0.017 & ( 0.948) &        &         &       &         \\
$\Delta_{BD_{1-2}}$ &-0.140 & (1.067) & 8.639 & (5.069) & 8.079 & ( 5.444) &        &         &       &         \\
$\Delta_{BD_{1-3}}$ &       &         &       &         & 8.817 & (13.149) &        &         &       &         \\
$\Delta_{BD_{1-4}}$ &       &         &       &         & 9.196 & ( 5.430) &        &         &       &         \\
     $p_{BD_{1-2}}$ & 0.991 &         & 0.178 &         & 0.491 &          &        &         &       &         \\
     $p_{BD_{1-3}}$ &       &         &       &         & 0.952 &          &        &         &       &         \\
     $p_{BD_{1-4}}$ &       &         &       &         & 0.355 &          &        &         &       &         \\
    \midrule
           $CD_1$ & 0.133 & (0.594) & 8.523 & (5.011) & 8.680 & ( 5.337) &  0.095 & (0.778) & 0.924 & (1.553) \\
           $CD_2$ & 0.427 & (0.680) & 0.063 & (0.460) & 1.250 & ( 1.438) &  6.539 & (3.198) & 1.606 & (1.081) \\
           $CD_3$ &       &         &       &         & 0.204 & ( 0.771) &        &         &       &         \\
           $CD_4$ &       &         &       &         &-0.345 & ( 0.714) &        &         &       &         \\
$\Delta_{CD_{1-2}}$ &-0.294 & (0.902) & 8.459 & (5.033) & 7.430 & ( 5.526) & -6.443 & (3.287) &-0.682 & (1.893) \\
$\Delta_{CD_{1-3}}$ &       &         &       &         & 8.476 & ( 5.385) &        &         &       &         \\
$\Delta_{CD_{1-4}}$ &       &         &       &         & 9.025 & ( 5.378) &        &         &       &         \\
     $p_{CD_{1-2}}$ & 0.943 &         & 0.186 &         & 0.588 &          &  0.105 &         & 0.934 &         \\
     $p_{CD_{1-3}}$ &       &         &       &         & 0.430 &          &        &         &       &         \\
     $p_{CD_{1-4}}$ &       &         &       &         & 0.365 &          &        &         &       &         \\
    \midrule
         Global p & 0.947 &         & 0.234 &         & 0.700 &          &  0.274 &         & 0.733 &         \\
         DIC      &98.843 &         &95.420 &         &96.354 &          & 95.745 &         &98.351 &    \\     
    \midrule
    \bottomrule
  \end{tabular}
  \bigskip
\end{table}

\begin{table}[!p]
  \caption{Results from ``leave-n-out'' node-split models for the
    Polish HIV data. $p_{U}$ denotes the unadjusted conflict p-value;
    $p_{AW}$ is the p-value adjusted for the multiple tests carried
    out \emph{within} each model (A)-(J) for the leave-2-out approach;
    $p_{AL}$ is the p-value adjusted for the 23 tests carried
    out in all models (A)-(J) for the leave-2-out approach; and
    $p_{AA}$ is the p-value adjusted for 28 tests carried out in all
    leave-1-out models (A)-(E) and all leave-2-out models (A)-(J). \label{tab_plResultsLnO}}
  \centering
  \begin{tabular}{ccccrrrr}
    \toprule
    \midrule
    Model & Partition 1 & Partition 2 & Node split & $p_U$ & $p_{AW}$
    & $p_{AL}$ & $p_{AA}$ \\
    \midrule
    \multicolumn{8}{c}{Leave-1-out} \\
    \midrule
    (A) & $y_1$ & $\{y_2,y_3,y_4,y_5\}$ & $\rho$         & $<0.0001$ &
    & 0.0060 & 0.0311\\
    (B) & $y_2$ & $\{y_1,y_3,y_4,y_5\}$ & $\pi\kappa$     & $<0.0001$
    & & 0.0047 & 0.0246\\
    (C) & $y_3$ & $\{y_1,y_2,y_4,y_5\}$ & $\pi(1-\kappa)$ & $0.6201$ &
    & 0.9857 & 1.0000\\
    (D) & $y_4$ & $\{y_1,y_2,y_3,y_5\}$ & $D_L$           & $0.1257$ &
    & 0.6242 & 0.9934\\
    (E) & $y_5$ & $\{y_1,y_2,y_3,y_4\}$ & $D_U$           & $<0.0001$
    & & 0.5852 & 0.9890\\
    \midrule
    \multicolumn{8}{c}{Leave-2-out} \\
    \midrule
    (A) & $\{y_1,y_2\}$ & $\{y_3,y_4,y_5\}$ & $\rho$          &
    0.6972  &   0.7480 & 1.0000 & 1.0000 \\
        &              &                   & $\pi\kappa$     &
        0.2209  &   0.2230 & 0.9842 & 0.9937 \\
    (B) & $\{y_1,y_3\}$ & $\{y_2,y_4,y_5\}$ & $\rho$          &
    $<0.0001$ &   0.0023  & 0.0240 & 0.0294\\
        &              &                   & $\pi(1-\kappa)$ &
        0.4906  &   0.7717 & 1.0000 & 1.0000 \\
    (C) & $\{y_2,y_3\}$ & $\{y_1,y_4,y_5\}$ & $\pi\kappa$     &
    $<0.0001$ & $<0.0010$ & $<0.0010$ & $<0.0010$\\
        &              &                   & $\pi(1-\kappa)$ &   0.8322  &   0.9000 & 1.0000 & 1.0000 \\
        &              &                   & $\pi$           &   0.9921  &   0.9490 & 1.0000 & 1.0000 \\
        &              &                   & $\kappa$        &   0.3329  &   0.6700 & 0.9998 & 1.0000 \\
        \midrule
    (D) & $\{y_1,y_4\}$ & $\{y_2,y_3,y_5\}$ & $\rho$          &
    $<0.0001$ &   0.0779 & 0.5754 & 0.6499 \\
        &              &                   & $D_L$           &
        0.0783  &   0.2851 & 0.9705 & 0.9866 \\
    (E) & $\{y_1,y_5\}$ & $\{y_2,y_3,y_4\}$ & $\rho$          &
    0.0745  &   0.1260 & 0.7614 & 0.8271 \\
        &              &                   & $D_U$           &
        0.0026  &   0.0949 & 0.6543 & 0.7276 \\
        \midrule
    (F) & $\{y_2,y_4\}$ & $\{y_1,y_3,y_5\}$ & $\pi\kappa$     &   0.4682  &   0.9590 & 1.0000 & 1.0000 \\
        &              &                   & $D_L$           &
        0.0869  &   0.3000 & 0.9764 & 0.9898 \\
    (G) & $\{y_2,y_5\}$ & $\{y_1,y_3,y_4\}$ & $\pi\kappa$     &   0.4420  &   0.6690 & 1.0000 & 1.0000 \\
        &              &                   & $D_U$           &
        0.0137  &   0.1970 & 0.9044 & 0.9434 \\
        \midrule
    (H) & $\{y_3,y_4\}$ & $\{y_1,y_2,y_5\}$ & $\pi(1-\kappa)$ &
    0.1471  &   0.3330 & 0.9855 & 0.9944 \\
        &              &                   & $D_L$           &
        0.1328  &   0.3280 & 0.9844& 0.9938 \\
    (I) & $\{y_3,y_5\}$ & $\{y_1,y_2,y_4\}$ & $\pi(1-\kappa)$ &   0.5237  &   0.8100 & 1.0000 & 1.0000 \\
        &              &                   & $D_U$           &
        $<0.0001$ &   0.2850 & 0.9702 & 0.9864 \\
        \midrule
    (J) & $\{y_4,y_5\}$ & $\{y_1,y_2,y_3\}$ & $D_L$           &
    0.1958  &   0.5117 & 0.9933 & 0.9978 \\
        &              &                   & $D_U$           &
        $<0.0001$ &   0.3963 & 0.9706 & 0.9866 \\
        &              &                   & $D$             &
        $<0.0001$ &   0.0030 & 0.0213 & 0.0260 \\
    \midrule
    \bottomrule
  \end{tabular}
  \bigskip
\end{table}

\clearpage
\newpage

\appendix

\section*{Supplementary Material}

\renewcommand{\thesubsection}{\Alph{subsection}}
\renewcommand\thefigure{\thesubsection.\arabic{figure}}    
\setcounter{figure}{0}    
\renewcommand\thetable{\thesubsection.\arabic{table}}    
\setcounter{table}{0}

\subsection{Figures and Tables}

\begin{figure}[hb]
  \centering
  \includegraphics[width = \textwidth]{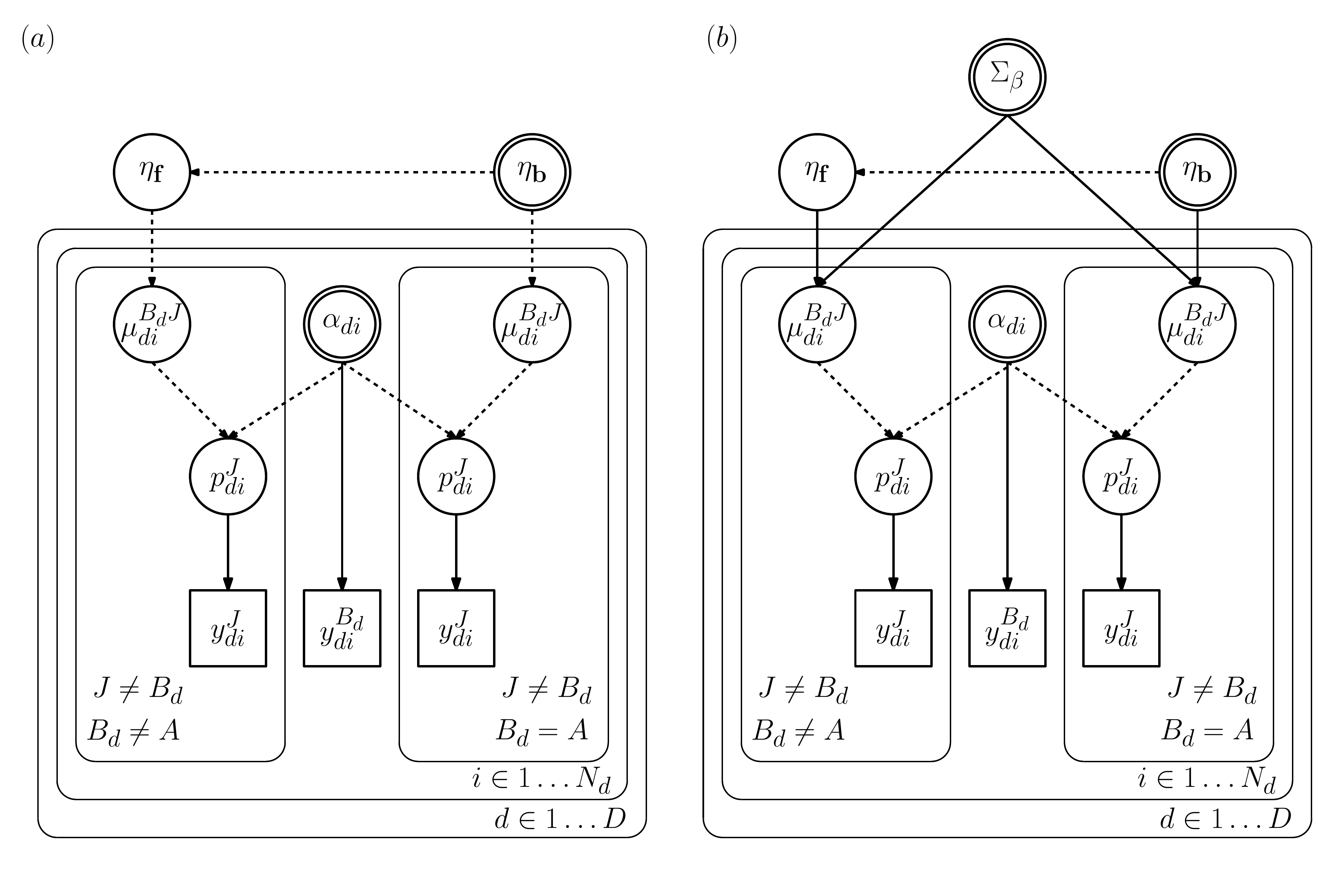}
  \caption{(a) DAG of NMA under assumptions of a common treatment effect $\eta^{JK}$ (no heterogeneity) and consistency $\eta^{JK} = \eta^{AK} - \eta^{AJ}$. (b) DAG of NMA under assumptions of random treatment effects, to account for heterogeneity, and consistency. \label{fig_nmaDAGs}}
\end{figure}

\begin{table}
  \caption{Smoking cessation data set}
  \label{tab_smkData}
  \centering
  \begin{tabular}{rlrrrrrrrrrrrr}
    \toprule
    \midrule
    Study & Design & $y_A$ & $n_A$ & $y_A/n_A$ & $y_B$ & $n_B$ & $y_B/n_B$ & $y_C$ & $n_C$ & $y_C/n_C$ & $y_D$ & $n_D$ & $y_D/n_D$ \\
    \midrule
    \midrule
     1 & AB  &   79 & 702 &0.113 &  77 & 694 &0.111 &   . &   . &    . &   . &   . &    . \\
     2 & AB  &   18 & 671 &0.027 &  21 & 535 &0.039 &   . &   . &    . &   . &   . &    . \\
     3 & AB  &    8 & 116 &0.069 &  19 & 149 &0.128 &   . &   . &    . &   . &   . &    . \\
    \midrule
     4 & AC  &   75 & 731 &0.103 &   . &   . &    . & 363 & 714 &0.508 &   . &   . &    . \\
     5 & AC  &    2 & 106 &0.019 &   . &   . &    . &   9 & 205 &0.044 &   . &   . &    . \\
     6 & AC  &   58 & 549 &0.106 &   . &   . &    . & 237 &1561 &0.152 &   . &   . &    . \\
     7 & AC  &    0 &  33 &0.000 &   . &   . &    . &   9 &  48 &0.188 &   . &   . &    . \\
     8 & AC  &    3 & 100 &0.030 &   . &   . &    . &  31 &  98 &0.316 &   . &   . &    . \\
     9 & AC  &    1 &  31 &0.032 &   . &   . &    . &  26 &  95 &0.274 &   . &   . &    . \\
    10 & AC  &    6 &  39 &0.154 &   . &   . &    . &  17 &  77 &0.221 &   . &   . &    . \\
    11 & AC  &   64 & 642 &0.100 &   . &   . &    . & 107 & 761 &0.141 &   . &   . &    . \\
    12 & AC  &    5 &  62 &0.081 &   . &   . &    . &   8 &  90 &0.089 &   . &   . &    . \\
    13 & AC  &   20 & 234 &0.085 &   . &   . &    . &  34 & 237 &0.143 &   . &   . &    . \\
    14 & AC  &   95 &1107 &0.086 &   . &   . &    . & 143 &1031 &0.139 &   . &   . &    . \\
    15 & AC  &   15 & 187 &0.080 &   . &   . &    . &  36 & 504 &0.071 &   . &   . &    . \\
    16 & AC  &   78 & 584 &0.134 &   . &   . &    . &  73 & 675 &0.108 &   . &   . &    . \\
    17 & AC  &   69 &1177 &0.059 &   . &   . &    . &  54 & 888 &0.061 &   . &   . &    . \\
    \midrule
    18 & ACD &    9 & 140 &0.064 &   . &   . &    . &  23 & 140 &0.164 &  10 & 138 &0.072 \\
    19 & AD  &    0 &  20 &0.000 &   . &   . &    . &   . &   . &    . &   9 &  20 &0.450 \\
    \midrule
    20 & BC  &    . &   . &    . &  20 &  49 &0.408 &  16 &  43 &0.372 &   . &   . &    . \\
    21 & BCD &    . &   . &    . &  11 &  78 &0.141 &  12 &  85 &0.141 &  29 & 170 &0.171 \\
    22 & BD  &    . &   . &    . &   7 &  66 &0.106 &   . &   . &    . &  32 & 127 &0.252 \\
    \midrule
    23 & CD  &    . &   . &    . &   . &   . &    . &  12 &  76 &0.158 &  20 &  74 &0.270 \\
    24 & CD  &    . &   . &    . &   . &   . &    . &   9 &  55 &0.164 &   3 &  26 &0.115 \\
    \midrule
    \bottomrule
  \end{tabular}
  \bigskip
\end{table}

\begin{table}
  \caption{Treatment effect posterior estimates (mean (sd)) for the common- and random-effect models respectively, with deviance summaries: posterior mean deviance $\mathbb{E}_{\theta \mid y}(D)$; deviance evaluated at posterior means $D(\mathbb{E}_{\theta \mid y}\theta)$; effective number of parameters $p_D$; and deviance information criterion $DIC$.}
  \label{tab_nmaConsistent}
  \centering
  \begin{tabular}{lrrrr}
    \toprule
    \midrule
    Model: & \multicolumn{2}{c}{Common-effect} & \multicolumn{2}{c}{Random-effect} \\
    $\mu^{JK}$: & Posterior mean & Posterior sd       & Posterior mean & Posterior sd \\
    \midrule
    $AB$                               & 0.224 & (0.124) & 0.496 & (0.405) \\
    $AC$                               & 0.765 & (0.059) & 0.843 & (0.236) \\
    $AD$                               & 0.840 & (0.174) & 1.103 & (0.439) \\
    $BC$                               & 0.541 & (0.132) & 0.347 & (0.419) \\
    $BD$                               & 0.616 & (0.192) & 0.607 & (0.492) \\
    $CD$                               & 0.075 & (0.171) & 0.260 & (0.418) \\
    \midrule
    $\mathbb{E}_{\theta \mid y}(D)$       &   267 &         &    54 & \\
    $D(\mathbb{E}_{\theta \mid y}\theta)$ &   240 &         &    10 & \\
    $p_D$                              &    27 &         &    44 & \\
    $DIC$                              &   294 &         &    98 & \\
    \midrule
    \bottomrule
  \end{tabular}
  \bigskip
\end{table}

\begin{table}
  \caption{Results from initial HIV model: observations; posterior mean (sd) estimates; posterior mean deviance $\mathbb{E}_{\theta \mid y}(D)$; deviance evaluated at posterior means $D(\mathbb{E}_{\theta \mid y}\theta)$; effective number of parameters $p_D$; and deviance information criterion $DIC$.}
  \label{tab_plResults}
  \centering
  \begin{tabular}{crrrrrrrrr}
    \toprule
    \midrule
    Parameter & \multicolumn{3}{c}{Data} & \multicolumn{2}{c}{Estimates} & \multicolumn{4}{c}{Deviance summaries} \\
    $\theta$  & $y$ & $n$ & $y/n$        & $\hat{y}$ & $\hat{\theta}$    & $\mathbb{E}_{\theta \mid y}(D)$ & $D(\mathbb{E}_{\theta \mid y}\theta)$ & $p_D$ & $DIC$ \\
    \midrule
$\rho$          &   35 & 1536 & 0.023 &   14.6 ( 1.5) & 0.010 (0.001) & 21.0 & 20.7 & 0.4 & 21.4 \\
$\pi\kappa$     &  113 & 2840 & 0.040 &   92.5 ( 8.9) & 0.033 (0.003) &  5.5 &  4.4 & 1.1 &  6.5 \\
$\pi(1-\kappa)$ &  136 & 2725 & 0.050 &  136.7 (11.3) & 0.050 (0.004) &  1.0 &  0.0 & 1.0 &  2.0 \\
$D_L$           &  836 &      &       &  836.2 (28.9) &  836.2 (28.9) &  1.0 &  0.0 & 1.0 &  2.0 \\
$D_U$           & 5034 &      &       & 5054.3 (70.8) & 5054.4 (70.8) &  1.1 &  0.1 & 1.0 &  2.1 \\
    \midrule
Total           &      &      &       &               &               & 29.5 & 25.1 & 4.4 & 33.9 \\
    \midrule
    \bottomrule
  \end{tabular}
  \bigskip
\end{table}

\begin{figure}
  \centering
  \includegraphics[width = \textwidth]{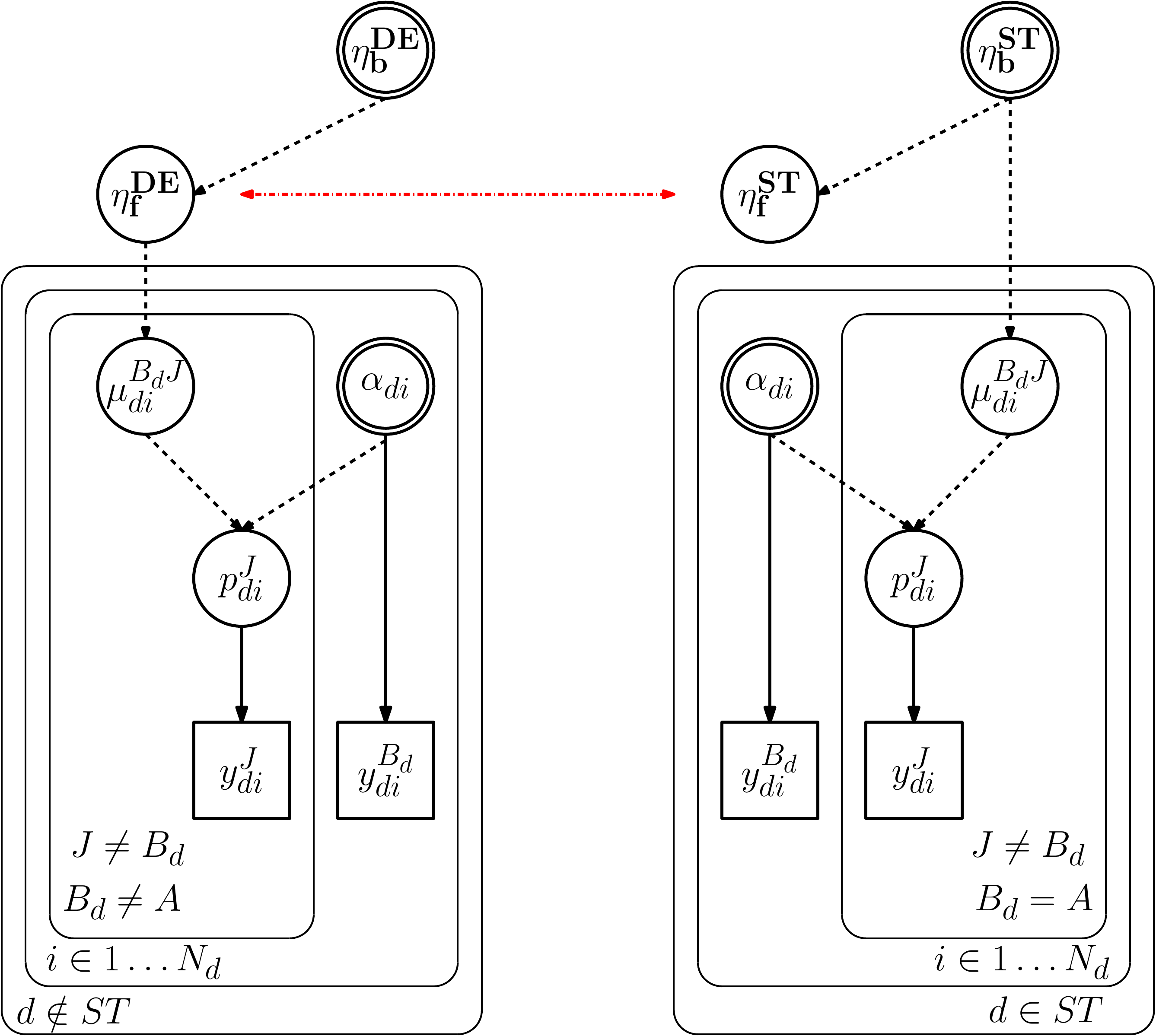}
  \caption{DAG of common-effect network meta-analysis model, split into direct (DE) and indirect (ST) evidence informing the functional parameters $\bs{\eta}_f$, i.e. those edges outside of the spanning tree formed by the basic parameters $\bs{\eta}_b$. \label{fig_nmaDAGsplit}}
\end{figure}

\begin{figure}
  \centering
  \includegraphics[width = 0.8\textwidth]{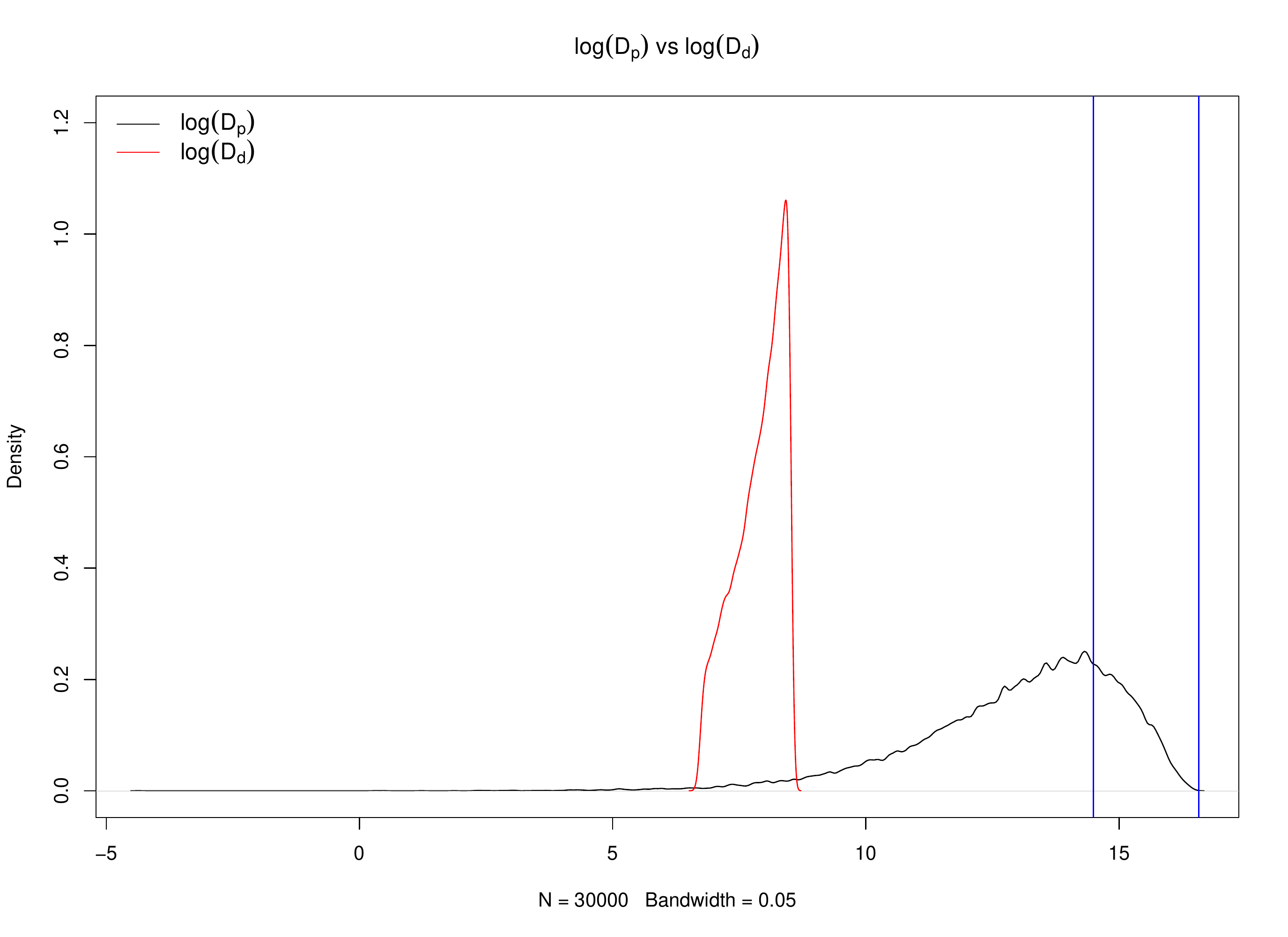}
  \includegraphics[width = 0.8\textwidth]{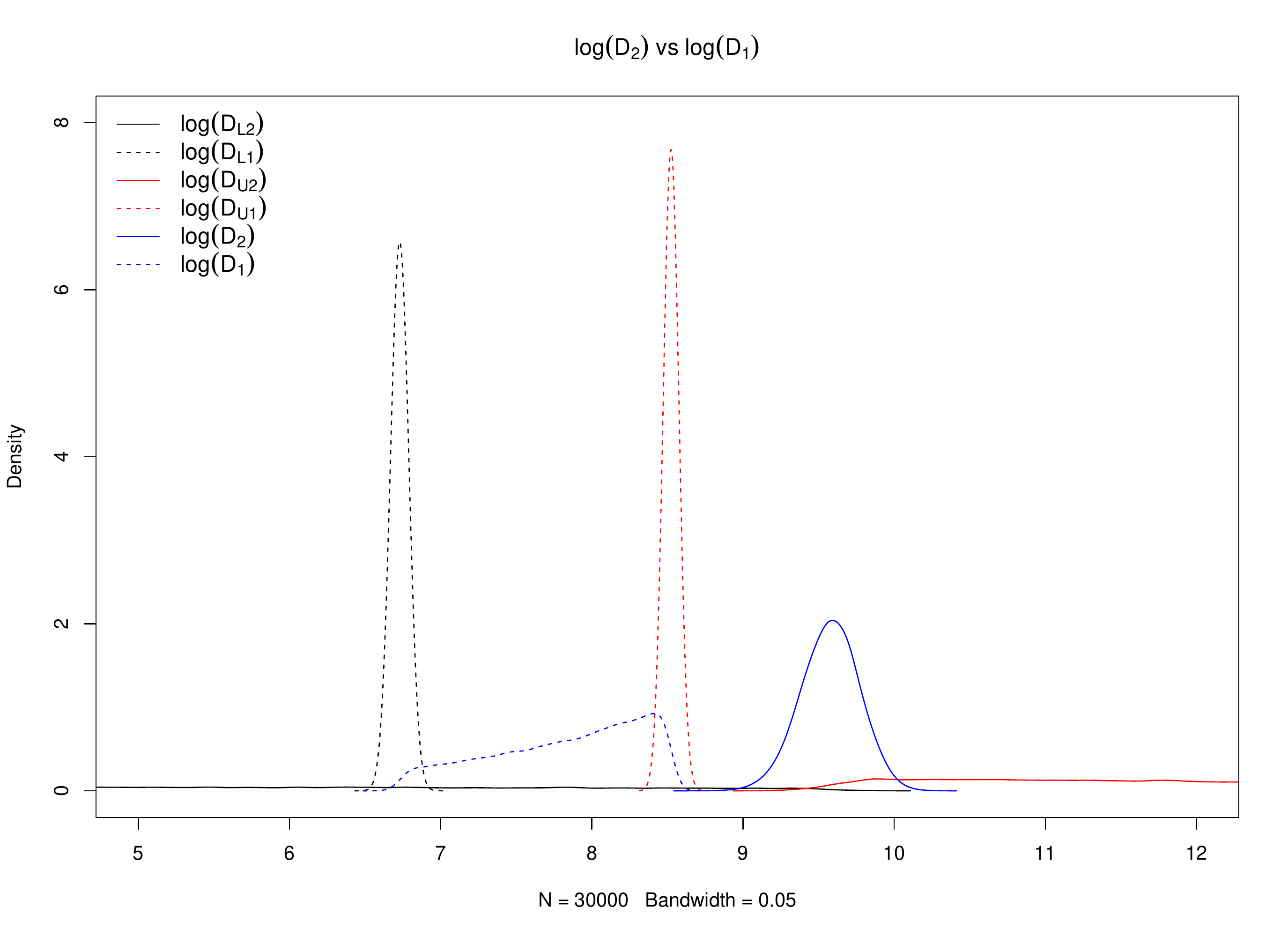}
  \caption{Upper panel: Posterior distributions of the nodes $D_p$ and $D_d$ for
    the HIV prevalence example, on the log scale. The right-hand blue
    line denotes where the total population of Poland ($N =
    15,749,944$) lies, i.e. the maximum possible value \emph{a priori}
    for the number diagnosed. The left-hand blue line denotes the
    value $\log(N \times 0.5^3)$, i.e. the prior mean of $\log(D_p) =
    \log(N\rho_p\pi_p\kappa_p)$. Lower panel: Posterior distributions
    of the nodes $D_{L1}, D_{L2}, D_{U1}, D_{U2}, D_1$ and $D_2$ for
    the HIV prevalence ``leave-2-out'' node-split model (J), on the
    log scale. The dashed lines represent the nodes in partition 1,
i.e. the ``left-out'' partition, where the posteriors are based only
on the likelihood given by $\{y_4,y_5\}$ and Jeffreys' priors for
$D_{L1}, D_{U1}$. The solid lines give the corresponding posteriors in
partition 2, i.e. based on all the original model priors and on the
dataset $\{y_1,y_2,y_3\}$. \label{fig_plSatD}}
\end{figure}

\clearpage

\subsection{Asymptotics \label{sec_AppBasymp}}
Let $p(\bs{\theta}_1), \ldots, p(\bs{\theta}_Q)$ denote the set of
prior distributions for the basic parameters $\bs{\theta}_q$ in each
partition $q$. Then by the independence of each partition, the
  joint posterior distribution of all parameters $\bs{\phi}$ in all
  partitions is
$$
p(\bs{\phi} \mid \bs{y}) = \prod_{q = 1}^Q p(\bs{\theta}_q) p(\bs{y}_q
\mid \bs{\theta}_q).
$$
If the joint prior distribution is dominated by the likelihood, 
then asymptotically \citep{BernardoSmith1994}, the joint posterior
distribution of all nodes 
is multi-variate normal:
$$
\bs{\phi} \mid \bs{y} \overset{a}{\sim} N_{\sum_q n_q} \left( (\bs{\hat{\phi}}_1, \ldots, \bs{\hat{\phi}}_Q), \bs{V} \right)
$$
where $n_q$ is the total number of parameters in partition $q$,
  whether basic or not, and $\bs{V}$ is the inverse observed
  information matrix for the parameters $\bs{\phi}$.
Since the vector of
  separator nodes, $\bs{\phi_S} = (\bs{\phi_1^{(s)}}, \ldots,
  \bs{\phi_Q^{(s)}})$, is a subset of $\bs{\phi}$, their joint
  posterior is also multivariate normal:
\begin{equation}
\bs{\phi_S} \mid \bs{y} \overset{a}{\sim} N_{m} \left( (\bs{\hat{\phi}}_1^{(s)}, \ldots, \bs{\hat{\phi}}_Q^{(s)}), \bs{V_S} \right) \label{eqn_bayes_asym}
\end{equation}
where $m = \sum_q m_q$ is the total number of separator nodes,
including node-split copies, and $\bs{V_S}$ is the appropriate
sub-matrix of $\bs{V}$. Since the partitions are independent,
$\bs{V_S}$ is a blocked diagonal matrix consisting of the inverse
observed information matrices for separator nodes in each partition along the diagonal.

By theorem 5.17 of \citet{BernardoSmith1994}, since
\eqref{eqn_bayes_asym} holds and if
$\bs{J_h}(\bs{\phi_S}) =
  \frac{\partial\bs{h}(\bs{\phi_S})}{\partial\bs{\phi_S}}$ is
non-singular with continuous entries, then the posterior distribution
of the transformed separator nodes, $\bs{\phi_H} = \bs{h}(\bs{\phi_S})$,
is also asymptotically normal:
$$
\bs{\phi_H} \mid \bs{y} \asym N_m \left(
  \bs{h}(\bs{\hat{\phi}}_1^{(s)}, \ldots, \bs{\hat{\phi}}_Q^{(s)}),
  \bs{J_h}(\bs{\hat{\phi}_S})^T \bs{V_S} \bs{J_h}(\bs{\hat{\phi}_S}) \right)
$$
The Jacobian $\bs{J_h}(\bs{\phi_S})$ exists and is non-singular for the sorts of
transformations we use in practice, for example log and logit
transformations.

A further application of theorem 5.17 of \citet{BernardoSmith1994}
results in a posterior distribution of the contrasts $\bs{\Delta}$
that is also aymptotically multivariate normal, if 
$\frac{\partial\bs{\Delta(\phi)}}{\partial\bs{\phi}} =
\bs{C_{\Delta}}^T$ is non-singular with continuous entries, which as a
contrast matrix it is:
\begin{eqnarray}
\bs{\Delta} \mid \bs{y} & \asym & N_p \left( \bs{C_{\Delta}}^T
  \bs{h}(\bs{\hat{\phi}}_1^{(s)}, \ldots, \bs{\hat{\phi}}_Q^{(s)}),
  \bs{C_{\Delta}}^T \bs{J_h}(\bs{\hat{\phi}_S})^T \bs{V_S}
  \bs{J_h}(\bs{\hat{\phi}_S}) \bs{C_{\Delta}}
\right) \label{eqn_bayes_asym_lincom_full} \\
 & = & N_p \left( \bs{C_{\Delta}}^T \hat{\bs{\phi}}_H,
   \bs{C_{\Delta}}^T \bs{V_H} \bs{C_{\Delta}}\right) \nonumber
\end{eqnarray}
for $\bs{V_H} = \bs{J_h}(\bs{\hat{\phi}_S})^T \bs{V_S}
  \bs{J_h}(\bs{\hat{\phi}_S})$. Asymptotically, therefore, the posterior mean $\overline{\bs{\Delta}} = \bs{C_{\Delta}}^T \overline{\bs{\phi_H}} \overset{a}{\approx} \bs{C_{\Delta}}^T \hat{\bs{\phi}}_H$ and the posterior variance-covariance matrix of $\bs{\Delta}$ is $\bs{S_{\Delta}} \overset{a}{\approx} \bs{C_{\Delta}}^T \bs{V_H} \bs{C_{\Delta}}$.

\end{document}